\documentclass[12pt]{article}
\usepackage{amsmath,amssymb,amsthm,mathrsfs,mathdots,mathtools}
\usepackage{graphicx,psfrag,epsf}
\usepackage[bookmarks,colorlinks]{hyperref}
\usepackage{bbm}
\usepackage{dsfont}
\usepackage{color}
\usepackage{enumerate}
\usepackage{natbib}
\usepackage{enumitem}
\usepackage{booktabs, arydshln}
\usepackage{url} 
\usepackage[labelformat=simple]{subfig}
\usepackage{dcolumn}
\usepackage{floatrow}
\usepackage{lscape}
\usepackage{float}

\newcommand{\blind}{0}

\hypersetup{colorlinks,%
citecolor=blue,%
filecolor=green,%
linkcolor=red,%
urlcolor=magenta,%
}

\theoremstyle{definition}
\newtheorem{defi}{Definition}[section]

\theoremstyle{definition}
\newtheorem{cond}{Condition}

\theoremstyle{plain}
\newtheorem{Theo}[defi]{Theorem}

\newtheorem{Coro}[defi]{Corollary}

\theoremstyle{definition}

\newcommand{\bfR}{{\boldsymbol{R}}}

\newcommand{\diff}{\mathrm{d}}

\newcommand{\surv}{\overline{F}}

\newcommand{\tod}{\stackrel{d}{\longrightarrow}}

\newcommand{\ind}{\mathbbm{1}}
\newcommand{\EE}{\mathbb E}
\newcommand{\PP}{\mathbb P}
\newcommand{\RR}{\mathbb R}

\newcommand{\Var}{\operatorname{Var}}
\DeclareMathOperator*{\argmin}{arg\,min}

\newlist{inparaenum}{enumerate}{2}
\setlist[inparaenum,1]{label=(\roman*)}
\setlist[inparaenum,2]{label=(\roman{inparaenumi}\emph{\alph*})}

\definecolor{navy}{rgb}{0,0,0.502}
\definecolor{brown}{rgb}{0.59, 0.29, 0.0}
\definecolor{green}{rgb}{0.25,0.50,0.25}

\begin{document}

\def\spacingset#1{\renewcommand{\baselinestretch}%
{#1}\small\normalsize} \spacingset{1}


\if0\blind
{
  \title{\bf Tail risk inference via expectiles in heavy-tailed time series}
  \author{A. C. Davison\hspace{.2cm}\\
    Institute of Mathematics, Ecole Polytechnique F\'ed\'erale de Lausanne\\
    S. A. Padoan\hspace{.2cm}\\
    Department of Decision Sciences, Bocconi University of Milan\\
    and \\
    G. Stupfler\hspace{.2cm}\\
    Universit\'e de Rennes, Ensai, CNRS, CREST}
  \maketitle
} \fi

\if1\blind
{
  \bigskip
  \bigskip
  \bigskip
  \begin{center}
    {\LARGE\bf Title}
\end{center}
  \medskip
} \fi

\bigskip
\begin{abstract}
Expectiles define the only law-invariant, coherent and elicitable risk measure apart from the expectation. The popularity of expectile-based risk measures is steadily growing and their properties have been studied for independent data, but further results are needed to use extreme expectiles with dependent time series such as financial data. In this paper we establish a basis for inference on extreme expectiles and expectile-based marginal expected shortfall in a general $\beta$-mixing context that encompasses ARMA, ARCH and GARCH models with heavy-tailed innovations. Simulations and applications to financial returns show that the new estimators and confidence intervals greatly improve on existing ones when the data are dependent.  
\end{abstract}

\noindent%
{\it Keywords:} Asymmetric least squares estimation, Marginal Expected Shortfall, Mixing, Tail copula, Weak dependence.
\vfill

\newpage
\spacingset{1.45} 

%
\section{Introduction}\label{sec:intro}
%
The quantification of the risk associated to a real-valued profit-loss variable $Y$ is a key problem in financial econometrics. The best-known financial risk measure, the Value-at-Risk, has serious deficiencies, as it does not generally define a coherent risk measure and thus does not obey the diversification principle~\citep{artdelebehea1999}, and it depends only on the frequency of tail losses and not on their actual values. A related measure, the expected shortfall, is coherent and takes into account the values of the risk variable, but it is not elicitable~\citep{gne2011} and thus does not benefit from the existence of a natural backtesting methodology.

A risk measure that addresses these issues is based on the  expectile~\citep{newpow1987}, which extends the mean in the same way that a quantile extends the median,  and is defined by minimising an asymmetrically weighted mean squared deviation,
\begin{equation}
\label{eq:expectile}
\xi_{\tau} = \argmin_{\theta\in \RR} \EE\{\eta_{\tau}(Y-\theta)-\eta_{\tau}(Y)\}, 
\end{equation}
where $Y$ is a real-valued random variable, $\EE$ denotes expectation, $\eta_{\tau}(u)=|\ind( u\leq 0 )-\tau| u^2$ with $\tau\in(0,1)$ determining the level of the expectile, and $\ind( \cdot )$ denotes the indicator function. The expectile is related to the gain-loss ratio, which is a popular performance measure in portfolio management~\citep[see][and references therein]{beldib2017}, and it depends on both the realisations of $Y$ and on their probabilities. Moreover, apart from the expectation it is the only risk measure that is law-invariant, coherent and elicitable.
Results of~\cite{beldib2017},~\citet{ehmgnejorkru2016} and others indicate that expectiles define sensible alternatives to the Value-at-Risk and expected shortfall, and inference for them has recently burgeoned; see \cite{holkla2016} and~\cite{krazah2017}, for example.  The estimation of extreme expectiles, for which $\tau\approx 1$,  was considered by~\cite{daogirstu2018,daogirstu2020}, but for independent and identically distributed data. Financial data are typically dependent, however, so these existing methods cannot safely be applied. Available techniques for expectile estimation with dependent data, such as those in~\cite{kuayehhsu2009},~\cite{xiezhowan2014} and~\cite{jiahuyu2020}, are restricted to central expectiles whose level stays well away from 0 and 1, so extreme expectile estimation for dependent data requires new tools.

In this paper we consider inference for extreme expectile-based risk measures based on $\beta$-mixing heavy-tailed strictly stationary time series, which encompass some widely-used time series models. 
To the best of our knowledge, we are the first to develop inference for extreme marginal expectiles in a time series context.
We start by considering the estimation of a tail expectile at the intermediate order, for which $\tau_n\to1$ such that $n(1-\tau_n)\to\infty$ as $n\to\infty$. Although a large expectile, $\xi_{\tau_n}$ is expected to typically fall within the range of the data.    
We then discuss the prediction of a tail expectile of order $\tau_n=1-p_n \to 1$ as the sample size $n\to\infty$ such that $n p_n$ is bounded; typically $p_n\leq 1/n$, and then the expectile is expected to fall in a neighbourhood of or above the largest observations available. We consider the least asymmetrically weighted squares estimator, which is the empirical counterpart of~\eqref{eq:expectile}, and a quantile-based estimator obtained from an asymptotic relationship between high expectiles and high quantiles. We then expand our statistical model to bivariate $\beta$-mixing time series $\{(X_t,Y_t), t=1,2,\ldots\}$ and establish estimation results for the marginal expected shortfall, which is important in assessing systemic risk~\citep{achpedphiric2017,broeng2017}. In this context, $X=X_1$ and $Y=Y_1$ represent the marginal losses on the returns of a company and of the entire market, and the marginal expected shortfall (MES) is $\EE(X \mid Y>u)$, where $u$ is large to reflect a substantial market downturn. Our results apply both to the expectile-based form of the extreme marginal expected shortfall, where $u$ is taken to be a high expectile of $Y$, and to its quantile-based counterpart. We discuss the construction of confidence intervals for extreme expectile-based risk measures and compare them with intervals obtained by assuming independent observations. 

Our methods and data have been incorporated into the {\tt R} package {\tt ExtremeRisks}, 
freely available on CRAN.  The online Supplementary Material contains technical results, further discussion and additional numerical work.
The layout of the paper is the following. Section~\ref{sec:framework} explains in detail our statistical context.  
Section~\ref{sec:expectiles} considers intermediate and extreme expectile estimation. Section~\ref{sec:MES} introduces a more general bivariate time series context within which general MES estimators at extreme levels are investigated. Section~\ref{sec:CI} provides a finite sample procedure for constructing confidence intervals  and discusses the selection of the expectile level in practice. The finite-sample performance of the methods is examined on simulated data in Section~\ref{sec:simulations_time_series} and on real financial data in Section~\ref{sec:real_data_time_series}.

%
%
\section{Statistical model and time series framework}\label{sec:framework} 

Let $(Y_t, t=1,2,\ldots)$ be a strictly stationary time series whose continuous marginal distribution $F$ has survival function $\surv=1-F$ and tail quantile function $U:s\mapsto \inf\{ y\in \RR \, : \, 1/\surv(y)\geq s\}$. Throughout the paper, $Y$ should be seen as the negative of a generic financial position, so large positive values of $Y$ represent extreme losses associated to this position.   Our goal is to estimate extreme expectiles of $Y$, so we focus on heavy-tailed distributions, which are ubiquitous in the modelling of extreme actuarial and financial losses~\citep{embklumik1997}. 
We therefore assume  for any $y>0$ that  as $s\to\infty$, $\surv(sy)/\surv(s)\to y^{-1/\gamma}$ or equivalently $U(sy)/U(s) \to y^{\gamma}$. 
The tail index $\gamma>0$ specifies the tail weight of $\surv$, whose tail becomes heavier as $\gamma$ increases. Together with the condition $\EE(|Y_-|)<\infty$, where $Y_-=\min(Y,0)$, the assumption $\gamma<1$ ensures that the first moment of $Y$ exists, and then the expectile $\xi_{\tau}$ is well-defined by~\eqref{eq:expectile}  for any $\tau$. 

Existing results on extreme expectile estimation require independent data~\citep{daogirstu2018}. Here we instead suppose that the time series is $\beta$-mixing.  For any $m=1,2,\ldots$, let $\mathcal{F}_{1,m}= \sigma(Y_1,\ldots,Y_m)$ and $\mathcal{F}_{m,\infty}= \sigma(Y_m, Y_{m+1},\ldots)$ denote the past and future $\sigma$-fields generated by $(Y_t)$. Then $(Y_t)$ is said to be $\beta$-mixing if the coefficients $\beta(l) = \sup_{m\geq 1} \beta_m(l)$ satisfy  $\beta(l) \to 0$ as $l\to \infty$, where
\[
\beta_m(l) = \EE [ \sup \{ |\PP(B\mid \mathcal{F}_{1,m}) - \PP(B)|: B\in \mathcal{F}_{m+l,\infty} \} ], \quad l=1,2,\ldots.
\]
\citet[][Section~2.4]{dou1994} shows that many Markov processes, including ARMA processes, nonlinear autoregressive processes and GARCH models, are geometrically $\beta$-mixing: under mild conditions there exists $a\in(0,1)$ such that $\beta(l) \leq a^l$ for  large enough $l$. The good balance between theoretical applicability and modelling strength offered by $\beta$-mixing motivates the following basic assumption. 

\begin{cond}\label{cond:model}
The time series $(Y_t,t=1,2\ldots)$ is strictly stationary and $\beta$-mixing  and its one-dimensional marginal distribution function $F$ is continuous and heavy-tailed.
\end{cond}

We henceforth suppose that our observations $Y_1,\ldots,Y_n$ are from a time series satisfying Condition~\ref{cond:model} and aim to estimate an unconditional marginal extreme expectile $\xi_{\tau'_n}$ of $Y=Y_1$, where $\tau'_n\to 1$ as $n\to\infty$.  The initial step is to estimate extreme but intermediate, i.e., in-sample, expectiles $\xi_{\tau_n}$, which we now discuss.

%
%
\section{Extreme expectile estimation in time series}\label{sec:expectiles} 
%

%
%
\subsection{Background}\label{sec:expectiles:inter}
%

Let $\tau_n$ represent a sequence of probabilities such that $\tau_n\to 1$ and $n(1-\tau_n)\to \infty$ as $n\to\infty$. A natural estimator of the expectile $\xi_{\tau_n}$ of the marginal distribution $F$ is the empirical counterpart of~\eqref{eq:expectile}, i.e., 
$$
\widetilde{\xi}_{\tau_n}=\argmin_{\theta\in \RR} \sum_{t=1}^n \eta_{\tau_n}(Y_t-\theta);
$$
this is readily computed, for example by iterative reweighted least squares. An alternative estimator exploits an asymptotic proportionality relationship between the high expectile $\xi_\tau$  and the corresponding quantile $q_\tau$: within our heavy-tailed model~\citep{belklamulgia2014},
\begin{equation}
\label{eq:proportionality}
\lim_{\tau\to 1}\frac{\xi_{\tau}}{q_{\tau}} = (\gamma^{-1}-1)^{-\gamma}, \quad \gamma<1.
\end{equation}
Thus a quantile-based estimator of $\xi_{\tau_n}$ is
$
\widehat{\xi}_{\tau_n} = (\widehat{\gamma}_n^{-1}-1)^{-\widehat{\gamma}_n} \widehat{q}_{\tau_n},
$
where $\widehat{q}_{\tau_n} = Y_{n-\lfloor n(1-\tau_n) \rfloor,n}$ is the empirical counterpart of $q_{\tau_n}$, $Y_{1,n}\leq \cdots\leq Y_{n,n}$ are the ascending order statistics of $Y_1,\ldots,Y_n$, and $\widehat{\gamma}_n$ is a consistent estimator of $\gamma$. Here $\lfloor x\rfloor$ is the floor function, that is, the largest integer smaller than or equal to $x$.
We focus on the following two simple examples. The first, the \citet{hil1975} estimator, 
\[
\widehat{\gamma}_n^H=\frac{1}{\lfloor n(1-\tau_n) \rfloor} \sum_{i=1}^{\lfloor n(1-\tau_n) \rfloor} \log\left(\frac{Y_{n-i+1,n}}{Y_{n-\lfloor n(1-\tau_n) \rfloor,n}}\right), 
\]
is the maximum likelihood estimator in a purely Pareto model, and thus is a natural estimator of $\gamma$. The second estimator exploits the relationship~\eqref{eq:proportionality}, which can be rephrased as $\overline{F}(\xi_{\tau})/(1-\tau) \to \gamma^{-1}-1$ as $\tau\uparrow 1$, which implies that 
\[
\gamma = \lim_{\tau\uparrow 1} \left( 1 + \frac{\overline{F}(\xi_{\tau})}{1-\tau} \right)^{-1}.
\]
Taking $\tau=\tau_n\to 1$, and estimating $\overline{F}(\xi_{\tau})$ by $\widehat{\overline{F}}_n(\widetilde{\xi}_{\tau_n})$, where $\widehat{\overline{F}}_n(u) = n^{-1} \sum_{t=1}^n \ind\{ Y_t>u \}$ is the empirical survival function, suggests the expectile-based (EB) estimator 
\[
\widehat{\gamma}_n^E = \left( 1 + \frac{\widehat{\overline{F}}_n(\widetilde{\xi}_{\tau_n})}{1-\tau_n} \right)^{-1}.
\]

Our first main results rely on the following conditions concerning dependence within the time series $(Y_t)$ and the gap between the right tail of $\overline{F}$ and a purely Pareto tail.

\begin{cond}\label{cond:time_series}
The time series $(Y_t,t=1,2,\ldots)$ has the following properties:
\begin{inparaenum}
\item \label{cond:enu:small_big_blocks} there exist sequences of integers $(l_n)$ and $(r_n)$ such that $l_n\to\infty$, $r_n\to\infty$, $l_n/r_n\to 0$, $ r_n/n\to 0$ and $n\,\beta(l_n)/r_n \to 0$ as $n\to\infty$; 

\item \label{cond:enu:tail_copula_time_series} for any $t= 1,2,\ldots$ there exists a function $R_t$ on $\mathcal{E}_2=[0,\infty]^2 \setminus \{  (\infty,\infty)\}$ such that 
\[
 \lim_{s\to \infty} s \, \PP\left\{ \overline{F}(Y_1)\leq \frac{x}{s}, \ \overline{F}(Y_{t+1})\leq \frac{y}{s} \right\} = R_t(x,y), \quad (x,y)\in \mathcal{E}_2; 
\]
\item \label{cond:enu:tail_copula_bound} there exist $D\geq 0$ and a nonnegative sequence $\rho(t)$ satisfying $\sum_{t= 1}^\infty \rho(t) <\infty$ and such that for $s$ large enough, any $t=1,2,\ldots$, and all $u,u',v,v'\in [0,1]$ such that $u'<u$ and $v'<v$, 
\begin{align*}
s\, \PP\!\left\{ \frac{u'}{s} < \overline{F}(Y_1)\leq \frac{u}{s}, \ \frac{v'}{s} < \overline{F}(Y_{t+1})\leq \frac{v}{s} \right\} \leq &\, \rho(t) \{(u-u')(v-v')\}^{1/2}\\ &\qquad + \frac{D}{s} (u-u')(v-v').
\end{align*}
\end{inparaenum}
\end{cond}
Conditions~\ref{cond:time_series}\ref{cond:enu:small_big_blocks} and~\ref{cond:time_series}\ref{cond:enu:tail_copula_time_series} have been employed previously in extreme value analysis with mixing conditions \citep[e.g.,][]{dre2003,dreroo2010}. The sequences $(l_n)$ and $(r_n)$ in Condition~\ref{cond:time_series}\ref{cond:enu:small_big_blocks} are small-block and big-block sequences used for standard arguments in the literature on mixing time series. Condition \ref{cond:time_series}\ref{cond:enu:tail_copula_bound} is slightly more precise than condition (C3) in~\cite{dre2003}.  Condition~\ref{cond:time_series} is discussed further in the Supplementary Material.

\begin{cond}\label{cond:RV2}
There exist $\rho\leq 0$ and a measurable function \textcolor{red}{$A(\cdot)$} having constant sign and converging to zero at infinity such that
\[
\lim_{s\to \infty}\frac{1}{A\{1/\overline{F}(s)\}} \left\{ \frac{\overline{F}(sy)}{\overline{F}(s)} - y^{-1/\gamma} \right\} = y^{-1/\gamma} \frac{y^{\rho/\gamma}-1}{\gamma\rho},\quad y>0; 
\]
 when $\rho=0$, the right-hand side should be read as $ \gamma^{-2}y^{-1/\gamma}\log y$. 
\end{cond}

\cite{beigoesegteu2004} and~\cite{haafer2006} give numerous examples of continuous distributions satisfying Condition~\ref{cond:RV2}.
We start by a general result on the least asymmetrically weighted squares estimator $\widetilde{\xi}_{\tau_n}$.
\begin{Theo}\label{theo:asydirintertime}
Assume that Conditions~\ref{cond:model} and~\ref{cond:time_series} are satisfied. Assume further that there exists a $\delta>0$ such that $\EE|Y_-|^{2+\delta}<\infty$, $0<\gamma<1/(2+\delta)$ and $\sum_{l\geq 1} [\beta(l)]^{\delta/(2+\delta)} <\infty$. 
Let $\tau_n\uparrow 1$ be such that $n(1-\tau_n)\to\infty$, $r_n(1-\tau_n)\to 0$ and $r_n \{r_n/\sqrt{n(1-\tau_n)}\}^{\delta}\to 0$ as $n\to\infty$. Then,
\[
\sqrt{n(1-\tau_n)} \left( \frac{\widetilde{\xi}_{\tau_n}}{\xi_{\tau_n}} - 1 \right) \tod \mathcal{N}\left( 0, \frac{2\gamma^3}{1-2\gamma} (1+\sigma^2(\gamma,\bfR)) \right),
\]
with 
$
\sigma^2(\gamma,\bfR) := \dfrac{(1-\gamma)(1-2\gamma)}{\gamma^2} \displaystyle\iint_{[1,\infty)^2} \sum_{t=1}^{\infty} R_t(x^{-1/\gamma},y^{-1/\gamma}) \, \diff x \, \diff y.
$
\end{Theo}
The family of functions $R_t$ specifies the extremal dependence within the time series between different time points;
when $R_t\equiv 0$ for any $t\geq 1$, which is for instance the case when $(Y_t)$ is an i.i.d.~sequence, the asymptotic variance is $2\gamma^3/(1-2\gamma)$.  
The quantity $\sigma^2(\gamma,\bfR)$ 
represents the proportion of increase of this asymptotic variance due to the mixing setting. The conditions $\EE|Y_-|^{2+\delta}<\infty$ and $0<\gamma<1/(2+\delta)$ essentially amount to assuming that the time series $Y_t$ has a finite variance.  
Theorem~\ref{theo:asydirintertime} above represents a substantial theoretical step compared to available results on extreme expectile estimation, because its proof is based on rather delicate arguments involving a tailored central limit theory for tail array sums in the time-dependent setting. 
Our assumptions are somewhat involved, but are very mild when $\beta(l)$ converges to 0 geometrically fast as $l\to\infty$. In that case, condition $\sum_{l\geq 1} [\beta(l)]^{\delta/(2+\delta)} <\infty$ is satisfied for any $\delta>0$, and one may choose $l_n=\lfloor C \log n\rfloor$, $r_n=\lfloor \log^2(n)\rfloor$ and $\tau_n=1-n^{-\tau}$, for any $\tau\in (0,1)$ and sufficiently large $C$. As we highlighted in Section~\ref{sec:framework}, geometrically strong $\beta$-mixing covers many widely-used financial time series models, including ARMA processes, ARCH/GARCH processes and solutions of stochastic difference equations~\citep[see][]{dou1994,dre2000,dre2003,frazak2006,boufucste2011}. The following corollary provides a rigorous statement in this geometrically mixing case.
\begin{Coro}\label{coro:asydirintertime}
Assume that Conditions~\ref{cond:model} and~\ref{cond:time_series}\ref{cond:enu:tail_copula_time_series}-\ref{cond:enu:tail_copula_bound} are satisfied, and that $\beta(l)=\operatorname{O}(a^l)$ for some $a\in(0,1)$. Assume further that there exists $\delta>0$ such that $\EE|Y|^{2+\delta}<\infty$. Let $\tau_n=1-n^{-\tau}$, for some $\tau\in (0,1)$. Then as $n\to\infty$, 
\[
\sqrt{n(1-\tau_n)} \left( \frac{\widetilde{\xi}_{\tau_n}}{\xi_{\tau_n}} - 1 \right) \tod \mathcal{N}\left( 0, \frac{2\gamma^3}{1-2\gamma} (1+\sigma^2(\gamma,\bfR)) \right),
\]
with the notation of Theorem~\ref{theo:asydirintertime}.
\end{Coro}
We turn to the analysis of the estimators $\widehat{\xi}_{\tau_n}$ based on $\widehat{\gamma}_n^{H}$ and $\widehat{\gamma}_n^E$. The crucial result in this case, which is of interest in its own right, consists in a joint Gaussian approximation of the processes $s\mapsto \widehat{q}_{1-(1-\tau_n)s} = Y_{n-\lfloor n(1-\tau_n)s \rfloor,n}$ and $s\mapsto \log \widehat{q}_{1-(1-\tau_n)s}$ (that is, the tail empirical quantile process and its logarithm) in our mixing framework. See also Theorem~2.4.8 in~\cite{haafer2006} for a result restricted to the independent case, as well as Proposition~A.1 in~\cite{haamerzho2016} and Proposition~1 in~\cite{chagui2018} for related but different statements that do not apply in our setup.
\begin{Theo}\label{theo:gaussapprox}
Assume that Conditions~\ref{cond:model},~\ref{cond:time_series} and~\ref{cond:RV2} are satisfied, and that $\tau_n\uparrow 1$, $n(1-\tau_n)\to\infty$, $r_n(1-\tau_n)\to 0$, $r_n \log^2(n(1-\tau_n))/\sqrt{n(1-\tau_n)} \to 0$ and $\sqrt{n(1-\tau_n)} A((1-\tau_n)^{-1}) = \operatorname{O}(1)$ as $n\to\infty$. Suppose finally that $n(1-\tau_n)$ is a sequence of integers and pick $s_0>0$. Then there exist appropriate versions of the process $s\mapsto \widehat{q}_{1-(1-\tau_n)s}$ and a continuous, centred Gaussian process $W$ having covariance function 
\[
r(x,y):=\min(x,y)+\sum_{t=1}^{\infty} R_t(x,y)+R_t(y,x)
\]
such that, for any $\varepsilon>0$ sufficiently small, we have, uniformly in $s\in (0,s_0]$,
\[
\frac{\widehat{q}_{1-(1-\tau_n)s}}{q_{\tau_n}} = s^{-\gamma} \left( 1 + \frac{1}{\sqrt{n(1-\tau_n)}} \gamma s^{-1} W(s) + \frac{s^{-\rho}-1}{\rho} A((1-\tau_n)^{-1}) + \operatorname{o}_{\PP}\left( \frac{s^{-1/2-\varepsilon}}{\sqrt{n(1-\tau_n)}} \right) \right)
\]
and 
\[
\log \frac{\widehat{q}_{1-(1-\tau_n)s}}{q_{\tau_n}} = -\gamma \log s + \frac{1}{\sqrt{n(1-\tau_n)}} \gamma s^{-1} W(s) + \frac{s^{-\rho}-1}{\rho} A((1-\tau_n)^{-1}) + \operatorname{o}_{\PP}\left( \frac{s^{-1/2-\varepsilon}}{\sqrt{n(1-\tau_n)}} \right).
\]
\end{Theo}
Condition~\ref{cond:time_series} is assumed in Theorem~\ref{theo:gaussapprox} for consistency with our framework; an inspection of the proof shows that Condition~\ref{cond:time_series}\ref{cond:enu:tail_copula_bound} can be replaced by its version with $u=u'=0$. As a consequence of Theorem~\ref{theo:gaussapprox}, one may determine the asymptotic behaviour of the pair $(\widehat{\gamma}_n^H,\widehat{q}_{\tau_n})$, which we then use in the following corollary to establish the limiting distribution of the estimator $\widehat{\xi}_{\tau_n}$ constructed using $\widehat{\gamma}_n^H$ as the tail index estimator. 
\begin{Coro}\label{coro:jtcovHillquant}
Assume that Conditions~\ref{cond:model},~\ref{cond:time_series} and~\ref{cond:RV2} are satisfied, with $\EE|Y_-|<\infty$ and $0<\gamma<1$. Let $\tau_n\uparrow 1$ be such that $n(1-\tau_n)\to\infty$, $r_n(1-\tau_n)\to 0$, $r_n \log^2(n(1-\tau_n))/\sqrt{n(1-\tau_n)} \to 0$, $\sqrt{n(1-\tau_n)} A((1-\tau_n)^{-1})\to \lambda_1\in \RR$ and $\sqrt{n(1-\tau_n)} q_{\tau_n}^{-1} \to \lambda_2\in \RR$ as $n\to\infty$. 
Then, for $\widehat{\gamma}_n=\widehat{\gamma}_n^H$ in the estimator $\widehat{\xi}_{\tau_n}$, one has
\[
\sqrt{n(1-\tau_n)} \left( \frac{\widehat{\xi}_{\tau_n}}{\xi_{\tau_n}} -1 \right)\tod \mathcal{N}\left( \frac{m(\gamma)}{1-\rho} \lambda_1 - \lambda, \gamma^2 \, v^H(\gamma,\bfR) \right)
\]
with 
\begin{align*}
\lambda &:= \left( \frac{(\gamma^{-1}-1)^{-\rho}}{1-\gamma-\rho} + \dfrac{(\gamma^{-1}-1)^{-\rho}-1}{\rho} \right) \lambda_1 + \gamma(\gamma^{-1}-1)^{\gamma} \EE(Y)\lambda_2 \\
\mbox{and } \ v^H(\gamma,\bfR) 	&:= (1+[m(\gamma)]^2) \left( 1+2\sum_{t=1}^{\infty} R_t(1,1) \right) \\[5pt]
								&+ 2 m(\gamma) \int_0^1 \sum_{t=1}^{\infty} \left[ \frac{R_t(s,1)+R_t(1,s)}{s} - 2R_t(1,1) \right] \diff s.
\end{align*}
\end{Coro}
Corollary~\ref{coro:jtcovHillquant} is, to the best of our knowledge, the first result on the estimator $\widehat{\xi}_{\tau_n}$ at intermediate levels under weak dependence assumptions (i.e., in the $\beta$-mixing framework). The asymptotic properties of $\widehat{\xi}_{\tau_n}$ using the estimator $\widehat{\gamma}_n^E$ are given in Theorem~C.5 in the Supplementary Material, where it can be seen that, due to asymptotic variance considerations, this estimator will tend to be less variable than the Hill estimator (in the i.i.d.~case, when $\gamma<0.38$). This may make it valuable in constructing confidence intervals requiring an estimate of $\gamma$, as we illustrate in Sections~B.1 and~B.2 of the Supplementary Material when constructing confidence intervals for intermediate expectiles.

%
\subsection{Extrapolation}\label{sec:expectiles:extreme}
%

We now consider estimating extreme expectiles $\xi_{\tau'_n}$ whose level $\tau'_n\to 1$ satisfies $n(1-\tau'_n)\to c\in [0,\infty)$ as $n\to\infty$. A typical choice in applications is $\tau'_n=1-p_n$ for an exceedance probability $p_n$ not greater than $1/n$ \citep[e.g.,][]{caieinhaazho2015}.
Our semiparametric approach  is motivated by a combination of the heavy-tailed assumption with Equation~\eqref{eq:proportionality} and defines an extreme expectile estimator through a \citet{wei1978}-type construction, whereby
\begin{equation}
\label{eq:approxW}
\frac{\xi_{\tau'_n}}{\xi_{\tau_n}} \approx \frac{q_{\tau'_n}}{q_{\tau_n}} = \frac{U\{(1-\tau'_n)^{-1}\}}{U\{(1-\tau_n)^{-1}\}} \approx \left( \frac{1-\tau'_n}{1-\tau_n} \right)^{-\gamma},\end{equation}
for $n$ large.  This suggests the following class of plug-in estimators of $\xi_{\tau'_n}$, 
\[
\overline{\xi}_{\tau'_n}^{\star}  \equiv \overline{\xi}_{\tau'_n}^{\star}(\tau_n) = \left( \frac{1-\tau'_n}{1-\tau_n} \right)^{-\widehat{\gamma}_n} \overline{\xi}_{\tau_n}, 
\]
where $\widehat{\gamma}_n$ and $\overline{\xi}_{\tau_n}$ are consistent estimators of $\gamma$ and of the expectile $\xi_{\tau_n}$.  We call $\overline{\xi}_{\tau'_n}^{\star}$ the extrapolating least asymmetrically weighted squares (LAWS) estimator when $\overline{\xi}_{\tau_n}=\widetilde{\xi}_{\tau_n}$, then denoting it by $\widetilde{\xi}_{\tau'_n}^{\star}$, and call it the extrapolating quantile-based (QB) estimator when $\overline{\xi}_{\tau_n}=\widehat{\xi}_{\tau_n}$, then denoting it by $\widehat{\xi}_{\tau'_n}^{\star}$. 
We can then prove the following result.
\begin{Theo}\label{theo:asyexttime}
Suppose that $\EE|Y_-|<\infty$ and that Conditions~\ref{cond:model},~\ref{cond:time_series} and~\ref{cond:RV2} are satisfied with $0<\gamma<1$ and $\rho<0$. Let $\tau_n,\tau'_n\uparrow 1$ with $n(1-\tau_n)\to \infty$, $n(1-\tau'_n)\to c\in [0,\infty)$ and $\{n(1-\tau_n)\}^{1/2}/\log\{(1-\tau_n)/(1-\tau'_n)\} \to \infty$ as $n\to\infty$. Assume also that $r_n(1-\tau_n)\to 0$, $r_n \{n(1-\tau_n)\}^{-1/2} \log^2\{n(1-\tau_n)\} \to 0$, $\{n(1-\tau_n)\}^{1/2} A\{(1-\tau_n)^{-1}\}\to \lambda_1\in \RR$ and $\{n(1-\tau_n)\}^{1/2} q_{\tau_n}^{-1} \to \lambda_2\in \RR$ as $n\to\infty$. Suppose finally that as $n\to\infty$, 
\[
\{n(1-\tau_n)\}^{1/2} \left( \frac{\overline{\xi}_{\tau_n}}{\xi_{\tau_n}}-1 \right) \to \Delta 
\]
weakly, where $\Delta$ is a nondegenerate random variable. Then, for $\widehat{\gamma}_n=\widehat{\gamma}_n^{H}$,
\[
\dfrac{\{n(1-\tau_n)\}^{1/2}}{\log\{(1-\tau_n)/(1-\tau_n')\}} \left( \dfrac{\overline{\xi}_{\tau_n'}^{\star}}{\xi_{\tau_n'}} - 1 \right) \to \mathcal{N}\left[ \dfrac{\lambda_1}{1-\rho}, \gamma^2 \left\{ 1+2\displaystyle\sum_{t=1}^{\infty} R_t(1,1) \right\} \right] 
\]
weakly as $n\to\infty$. 
\end{Theo}
Theorem~\ref{theo:asyexttime} enables the construction of confidence intervals for extreme expectiles. 
The quantity $2\sum_{t=1}^{\infty} R_t(1,1)$ represents the relative increase in the asymptotic variance due to the temporal dependence. When $R_t\equiv 0$ for any $t\geq 1$, the asymptotic variance is $\gamma^2$. An analogous result may be obtained for the estimator $\widehat{\gamma}_n^E$, although we shall not pursue this for the sake of brevity.

\section{Marginal expected shortfall estimation}
\label{sec:MES}

It is important to assess overall, or systemic, risk when working with actuarial and financial data, for instance by  simultaneously considering several lines of business of an insurance company or several stock market indices.  A prominent way to measure such risk is via the marginal expected shortfall, defined as the propensity of a financial institution to be undercapitalised when the financial system as a whole is undercapitalised~\citep{achpedphiric2017,broeng2017}. The contribution that an individual firm with loss return $X$ makes to systemic risk, represented by a loss $Y$ in the aggregated return of the market, can be measured by the quantile-based marginal expected shortfall 
\begin{equation}
\label{eq:trueqmes}
\textrm{QMES}_{X,\tau} = \EE(X\mid Y > q_{Y,\tau}), \quad  \tau\in (0,1),
\end{equation}
where $q_{Y,\tau}$ is the $\tau$ quantile of $Y$. A systemic crisis typically corresponds to a situation in which $\tau$ is close to or exceeds $1-1/n$, where $n$ is the historical sample size.  An alternative to~\eqref{eq:trueqmes}  is the expectile-based marginal expected shortfall, 
\begin{equation}
\label{eq:truexmes}
\textrm{XMES}_{X,\tau} = \EE(X\mid Y > \xi_{Y,\tau}), \quad \tau\in (0,1),
\end{equation}
with $\xi_{Y,\tau}$ the $\tau$ expectile of $Y$.  Estimation of $\textrm{QMES}_{X,\tau}$ and $\textrm{XMES}_{X,\tau}$ at extreme levels is considered by~\cite{caieinhaazho2015} and~\cite{daogirstu2018}, but only for independent data.  

We now extend the results of Section~\ref{sec:expectiles} to inference for these definitions of marginal expected shortfall at extreme levels in our weakly-dependent setting. Suppose that the data come from a strictly stationary bivariate time series $\{(X_t,Y_t), t=1,2,\ldots\}$; for instance, $X_t$ and $Y_t$ could be the daily loss returns on a specific stock and on a market index. For any $m\geq 1$, let $\mathcal{F}_{1,m}= \sigma(X_1,Y_1,\ldots,X_m,Y_m)$ and $\mathcal{F}_{m,\infty}= \sigma(X_m, Y_m,X_{m+1},Y_{m+1},\ldots)$ denote the past and future $\sigma$-fields generated by $(X_t,Y_t)$. Then the corresponding $\beta$-mixing coefficients  can be defined as $b(l)=\sup_{m\geq 1} b_m(l)$, where
\[
b_m(l) = \EE[ \sup \{ |\PP(B\mid \mathcal{F}_{1,m}) - \PP(B)|: B\in \mathcal{F}_{m+l,\infty} \} ], \quad l=1,2\ldots.
\]
The sequence $\{(X_t,Y_t), t=1,2,\ldots\}$ is then said to be $\beta$-mixing if $b(l)\to 0$ as $l\to\infty$. If it is $\beta$-mixing, then $(Y_t,t=1,2,\ldots)$ is also $\beta$-mixing in the sense of Section~\ref{sec:framework}. Our condition below 
similarly extends Conditions~\ref{cond:model} and~\ref{cond:time_series} in a natural way.
\begin{cond}
\label{cond:time_series_2D}
\begin{inparaenum}
\item \label{cond:enu:model_2D} The time series $\{(X_t,Y_t), t=1,2,\ldots\}$ is strictly stationary, $\beta$-mixing and the one-dimensional marginal distribution functions $F_X$ and $F_Y$ of $(X_t, t=1,2,\ldots)$ and $(Y_t, t=1,2,\ldots)$ are continuous and heavy-tailed with respective tail indices $\gamma_X$ and $\gamma_Y$; 
\item \label{cond:enu:small_big_blocks_2D} there exist sequences of integers $(l_n)$ and $(r_n)$ such that 
$l_n\to\infty$, $r_n\to\infty$, $l_n/r_n\to 0$, $r_n/n\to 0$ and $nb(l_n)/r_n \to 0$ as $n\to\infty$; 
\item \label{cond:enu:tail_copula_time_series_2D} for any $t=1,2,\ldots$ there exists a function $r_t$ on $\mathcal{E}_4=[0,\infty]^4 \setminus \{  (\infty,\infty,\infty,\infty) \}$ such that 
$$
 \lim_{s\to \infty} s \, \PP\left\{ \overline{F}_X(X_1)\leq \frac{x_1}{s}, \ \overline{F}_X(X_{t+1})\leq \frac{x_{t+1}}{s}, \ \overline{F}_Y(Y_1)\leq \frac{y_1}{s}, \ \overline{F}_Y(Y_{t+1})\leq \frac{y_{t+1}}{s} \right\} 
 $$
 equals $r_t(x_1,x_{t+1},y_1,y_{t+1}) $ for any $(x_1,x_{t+1},y_1,y_{t+1}) \in \mathcal{E}_4$; and  

\item \label{cond:enu:tail_copula_bound_2D} there exist $D\geq 0$ and a nonnegative sequence $\rho(t)$ satisfying $\sum_{t= 1}^\infty \rho(t) <\infty$ such that if $s$ is large enough, 
for any $t=1,2,\ldots$, all $x_1,x_{t+1}\in [0,\infty]$, and all $u,u',v,v'\in [0,1]$ with $u'<u$ and $v'<v$, 
$$
s \, \PP\left\{ \overline{F}_X(X_1)\leq \frac{x_1}{s}, \ \overline{F}_X(X_{t+1})\leq \frac{x_{t+1}}{s}, \ \frac{u'}{s} < \overline{F}_Y(Y_1)\leq \frac{u}{s}, \ \frac{v'}{s} < \overline{F}_Y(Y_{t+1})\leq \frac{v}{s} \right\} 
$$ 
is no greater than 
$$
\rho(t) \{\min(x_1,u-u') \min(x_{t+1},v-v')\}^{1/2} + \frac{D}{s} \min(x_1,u-u') \min(x_{t+1},v-v').
$$
\end{inparaenum}
\end{cond}
If $\{(X_t,Y_t), t=1,2,\ldots\}$ satisfies Condition~\ref{cond:time_series_2D}, then the univariate time series $(Y_t,t=1,2,\ldots)$ automatically satisfies Conditions~\ref{cond:model} and~\ref{cond:time_series}, since if we set $x_1=x_{t+1}=\infty$ then the function $R_t\equiv R_{Y,t}$ is given by $R_{Y,t}(y_1,y_{t+1}) = r_t(\infty,\infty,y_1,y_{t+1})$. 

We now embed~\eqref{eq:trueqmes} and~\eqref{eq:truexmes} in a more general marginal expected shortfall framework. Define 
$\textrm{MES}_{X,\tau} = \EE(X\mid Y > z_{Y,\tau})$ for $ \tau\in (0,1)$, where $z_{Y,\tau}$ is a risk measure on $Y$ such that $\overline{F}(z_{Y,\tau})/(1-\tau)\to z=z(\gamma_Y)\in (0,\infty)$ as $\tau\uparrow 1$. If an $\{n(1-\tau_n)\}^{1/2}$-relatively consistent estimator $\overline{z}_{Y,\tau_n}$ of $z_{Y,\tau_n}$ is available at the intermediate level $\tau_n$, then an extrapolation relationship similar to~\eqref{eq:approxW} suggests defining an estimator of $\textrm{MES}_{X,\tau'_n} = \EE(X\mid Y > z_{Y,\tau'_n})$ as 
\[
\overline{\textrm{MES}}_{X,\tau'_n}^{\star} \equiv \overline{\textrm{MES}}_{X,\tau'_n}^{\star}(\tau_n) = \left( \frac{1-\tau'_n}{1-\tau_n} \right)^{-\widehat{\gamma}_{X,n}} \frac{\sum_{t=1}^n X_t \,\ind( X_t>0, Y_t>\overline{z}_{Y,\tau_n} )}{\sum_{t=1}^n \ind( Y_t>\overline{z}_{Y,\tau_n} )}.
\]
Replacing $\overline{z}_{Y,\tau_n}$ by $\widehat{q}_{{\tau}_n}$, $\widetilde{\xi}_{\tau_n}$ or $\widehat{\xi}_{\tau_n}$ yields estimators $\widehat{\textrm{QMES}}_{X,\tau'_n}^{\star}$, $\widetilde{\textrm{XMES}}_{X,\tau'_n}^{\star}$ or $\widehat{\mathrm{XMES}}_{X,\tau'_n}^{\star}$ of~\eqref{eq:trueqmes} or~\eqref{eq:truexmes}. 
The following second-order condition is needed to quantify the bias incurred in constructing these estimators. Below we use the classical asymptotic notation $o(\cdot)$, $O(\cdot)$ and $O_{\PP}(\cdot)$ \citep{dodge2006oxford}.
\begin{cond}
\label{cond:JC2}
Under Conditions~\ref{cond:time_series_2D}\ref{cond:enu:model_2D} and \ref{cond:time_series_2D}\ref{cond:enu:tail_copula_time_series_2D}, suppose that there exist $\beta>\gamma_X$ and $\kappa<0$ such that $R_{(X,Y)}(x,y) = r_1(x,\infty,y,\infty)$ satisfies $R_{(X,Y)}(1,1)>0$ and
\[
\sup_{x\in (0,\infty)} \left| \frac{s \, \PP\{\overline{F}_X(X_1)\leq x/s, \overline{F}_Y(Y_1)\leq y/s\} - R_{(X,Y)}(x,y)}{\min(x^{\beta},1)} \right| = O(s^{\kappa}) 
\]
locally uniformly in $y\in (0,\infty)$ as $s\to\infty$. 
\end{cond}
%
The following theorem gives the asymptotic distribution of $\overline{\textrm{MES}}^{\star}_{X,\tau'_n}$.
\begin{Theo}\label{theo:asyMES} 
Suppose that $X=X_1$ and $Y=Y_1$ satisfy Condition~\ref{cond:RV2} with respective parameters $(\gamma_X,\rho_X,A_X)$ and $(\gamma_Y,\rho_Y,A_Y)$, and that Conditions~\ref{cond:time_series_2D} and~\ref{cond:JC2} hold. Suppose also that $\rho_X<0$, that there exists $\delta>0$ such that $0<\gamma_X<1/(2+\delta)$, and that
\begin{enumerate}[label=(\roman*)]
\item $\tau_n$, $\tau'_n\uparrow 1$, with $n(1-\tau_n)\to \infty$, $n(1-\tau'_n)\to c<\infty$ and $\{n(1-\tau_n)\}^{1/2}/\log\{(1-\tau_n)/(1-\tau'_n)\} \to \infty$ as $n\to \infty$; 
\item $r_n(1-\tau_n)\to 0$ and $r_n [r_n\{n(1-\tau_n)\}^{-1/2}]^{\delta}\to 0$ as $n\to\infty$; 
\item there exists  $\varepsilon>0$ such that $\{n(1-\tau_n)\}^{1/2} |A_X\{(1-\tau_n)^{-1}\}|^{\gamma_X/(1-\rho_X) - \varepsilon}\to 0$, and $n (1-\tau_n)^{1-2\kappa} \to 0$ as $n\to\infty$; 
\item $\EE(|X_-|^{1/\gamma_X})<\infty$ and $n(1-\tau_n)\times (1-\tau'_n)^{-2\kappa(1-\gamma_X)}\to 0$ as $n\to\infty$;
\item the bias conditions 
\[
\{n(1-\tau_n)\}^{1/2} \left\{ \frac{\overline{F}_Y(z_{Y,\tau_n})}{1-\tau_n} - z \right\} = o(1), \quad  \{n(1-\tau_n)\}^{1/2} \left\{ \frac{\overline{F}_Y(z_{Y,\tau'_n})}{1-\tau'_n} - z \right\} = o(1),
\]
hold as $n\to\infty$;  and
\item the weak convergence $\{n(1-\tau_n)\}^{1/2} (\widehat{\gamma}_{X,n}-\gamma_X) \to \Gamma$ holds, where $\Gamma$ is a nondegenerate random variable, and
\[
\{n(1-\tau_n)\}^{1/2} \left( \frac{\overline{z}_{Y,\tau_n}}{z_{Y,\tau_n}} - 1 \right) = O_{\PP}(1).
\]
\end{enumerate}
If also $\{n(1-\tau_n)\}^{1/2} A_Y\{(1-\tau_n)^{-1}\}\to 0$ as $n\to\infty$, then
$$
\frac{\{n(1-\tau_n)\}^{1/2}}{\log\{(1-\tau_n)/(1-\tau'_n)\}} \left( \frac{\overline{\mathrm{MES}}_{X,\tau'_n}^{\star}}{\mathrm{MES}_{X,\tau'_n}} -1 \right) \to \Gamma
$$
weakly. Condition (iv) is unnecessary if $X>0$ with probability one.
\end{Theo}
Condition~(i) was used in Theorem~\ref{theo:asyexttime}. Condition~(ii) is used to deal with serial dependence. Condition~(iii) is slightly weaker than condition~(d) of~\cite{caieinhaazho2015}, and Condition~(iv) is taken from Theorem~2 therein. Condition~(v) is used to control the error made in using the extrapolation relationship, and Condition~(vi) ensures that all quantities used in the construction of the estimators converge at the appropriate rate. In this condition, $\Gamma$ is typically a normal distribution, as is for instance the case when $\widehat{\gamma}_{X,n}$ is a Hill estimator.
%
%
Using Theorem~\ref{theo:asyMES}, we obtain the following important corollary.
\begin{Coro}
\label{coro:asyQMES} 
Under the conditions of Theorem~\ref{theo:asyMES} (apart from (v) and (vi)), assume also that $r_n \log^2(n(1-\tau_n))/\sqrt{n(1-\tau_n)} \to 0$ as $n\to\infty$ and let $\widehat{\gamma}_{X,n}=\widehat{\gamma}_{X,n}^H$. 
Let $R_{X,t}(x_1,x_{t+1}) := r_t(x_1,x_{t+1},\infty,\infty)$. Then as $n\to\infty$ the weak convergence 
\[
\frac{\sqrt{n(1-\tau_n)}}{\log\{(1-\tau_n)/(1-\tau'_n)\}} \left( \frac{\overline{\mathrm{MES}}_{X,\tau'_n}^{\star}}{\mathrm{MES}_{X,\tau'_n}} -1 \right) \to \mathcal{N}\left[ 0, \gamma_X^2 \left\{ 1+2\displaystyle\sum_{t=1}^{\infty} R_{X,t}(1,1) \right\} \right]
\]
holds for
\begin{itemize}
\item $\overline{\mathrm{MES}}_{X,\tau'_n}^{\star}=\widehat{\mathrm{QMES}}_{X,\tau'_n}^{\star}$, without further assumptions, 
\item $\overline{\mathrm{MES}}_{X,\tau'_n}^{\star}=\widetilde{\mathrm{XMES}}_{X,\tau'_n}^{\star}$, assuming further with the notation of Theorem~\ref{theo:asyMES} that $\EE|Y_-|^{2+\delta}<\infty$, $0<\gamma_Y<1/(2+\delta)$, $\sum_{l\geq 1} [b(l)]^{\delta/(2+\delta)} <\infty$ and that $\sqrt{n(1-\tau_n)} q_{Y,\tau_n}^{-1} \to 0$ as $n\to\infty$,  and
\item $\overline{\mathrm{MES}}_{X,\tau'_n}^{\star}=\widehat{\mathrm{XMES}}_{X,\tau'_n}^{\star}$, assuming further with the notation of Theorem~\ref{theo:asyMES} that $\EE|Y_-|<\infty$, $0<\gamma_Y<1$, and that $\sqrt{n(1-\tau_n)} q_{Y,\tau_n}^{-1} \to 0$ and $\sqrt{n(1-\tau_n)} (\widehat{\gamma}_{Y,n}-\gamma_Y) = \operatorname{O}_{\PP}(1)$ as $n\to\infty$.
\end{itemize}
\end{Coro}
%

\section{Finite-sample inference and expectile level selection}
\label{sec:CI} 
%
The theory developed in Sections~\ref{sec:expectiles} and~\ref{sec:MES} allows us to obtain confidence intervals for extreme expectiles and the expectile-based marginal expected shortfall, provided we can estimate the asymptotic variance $w(\gamma,\bfR)=\gamma^2 \{ 1+2\sum_{t=1}^{\infty} R_t(1,1) \}$ in Theorem~\ref{theo:asyexttime}. By Proposition 2.1 in \citet{dre2003} we have, when $n(1-\tau_n)\to\infty$, $r_n\to \infty$ and $r_n(1-\tau_n)\to 0$,
\[
\frac{1}{r_n(1-\tau_n)}\Var\left[ \sum_{i=1}^{r_n}\ind\{F(Y_i)>\tau_n\}\right] \to 1+2\displaystyle\sum_{t=1}^{\infty} R_t(1,1) ,\quad n\to\infty.
\]
We split the data into big blocks of length $r_n$ separated by small blocks of length $l_n$ and define 
$$
Z_j = \sum_{t=1+j\ell_n}^{r_n+j\ell_n} 1\{\widehat{F}_n(Y_t)>\tau_n\}, \quad j=0,1,\ldots,m_n-1, 
$$
where $m_n=\lfloor n/\ell_n\rfloor$, $\ell_n=r_n+l_n$, and $\widehat{F}_n$ is the empirical distribution function of all the data. We then compute the sample variance $\Sigma_n$ of $Z_0,\ldots,Z_{m_n-1}$ and thus obtain an estimator
\begin{equation}\label{eq:aymp_var_emp_est_extr}
\widehat{w}_n(\gamma,\bfR)=\frac{(\widehat{\gamma}_n^{H})^2}{r_n(1-\tau_n)}\Sigma_n
\end{equation}
of $w(\gamma,\bfR)$.  
Theorem~\ref{theo:asyexttime} yields as $n\to\infty$ the weak convergence 
\[
\frac{\{n(1-\tau_n)\}^{1/2}}{\log\{(1-\tau_n)/(1-\tau_n')\}} \log\frac{\overline{\xi}_{\tau_n'}^{\star}}{\xi_{\tau_n'}} \to \mathcal{N}\left\{ \frac{\lambda_1}{1-\rho}, w(\gamma,\bfR) \right\}.
\]
Thus a $(1-\alpha)$-equi-tailed confidence interval for ${\xi_{\tau_n'}}$ has limits 
\begin{equation}\label{eq:conf_int_weiss}
\overline{\xi}^{\star}_{\tau_n'}\left(\frac{1-\tau_n}{1-\tau_n'}\right)^{-\widehat{b}\pm z_{1-\alpha/2}[\widehat{w}_n(\gamma,\bfR)/\{n(1-\tau_n)\}]^{1/2}}, 
\end{equation}
where $\overline{\xi}_{\tau'_n}^{\star}$ is the least asymmetrically weighted squares or quantile-based extrapolating estimator, $\widehat{b} = \widehat{\gamma}_n^{H} \widehat{\beta}_n (1-\tau_n)^{-\widehat{\rho}_n}/(1-\widehat{\rho}_n)$ estimates the bias $\lambda_1/(1-\rho)$, with the estimates $\widehat{\beta}_n$ and $\widehat{\rho}_n$ obtained from the {\tt R} package {\tt evt0}~\citep{mancae2013}, and $z_{p}$ the $p$-quantile of the standard normal distribution. The intervals given by~\eqref{eq:conf_int_weiss} can be constructed with  $\overline{\xi}^{\star}_{\tau_n'}$ either the least asymmetrically weighted squares or the quantile-based extrapolating estimator, and ignoring the bias, i.e., setting  $\widehat b=0$.  The asymptotic variance in the independent case, $\gamma^2$, is estimated by $(\widehat{\gamma}_n^H)^2$.  Similar confidence intervals can be constructed for \eqref{eq:truexmes}, by substituting $X_t$ for $Y_t$ in both the $Z_j$ and the Hill estimator, and by replacing $\overline{\xi}^{\star}_{\tau_n'}$ either by $\widetilde{\mathrm{XMES}}_{X,\tau'_n}^{\star}$ or by $\widehat{\mathrm{XMES}}_{X,\tau'_n}^{\star}$.


A crucial issue in risk management is the choice of the prudentiality level $\tau'_n$. If expectiles are considered for their financial interpretation in terms of the gain-loss ratio, then $\tau'_n$ should be selected to achieve a certain gain-loss value; otherwise, it seems reasonable to select $\tau'_n$ so that $\xi_{\tau'_n} \equiv q_{\alpha_n}$ for a given $\alpha_n$. In our context this level $\tau'_n=\tau'_n(\alpha_n)$ satisfies
$
\{1-\tau_n'(\alpha_n)\}/(1-\alpha_n) \to \gamma/(1-\gamma) 
$
as $n\to\infty$. One can then define a natural estimator of $\tau_n'(\alpha_n)$, i.e., 
\[
\widehat{\tau}_n'(\alpha_n) = 1-(1-\alpha_n) \, \frac{\widehat{\gamma}_n^{H}}{1-\widehat{\gamma}_n^{H}}.
\]
Using this estimator in place of $\tau_n'$ in the extrapolating least asymmetrically weighted squares and quantile-based estimators yields composite estimators of $\xi_{\tau'_n(\alpha_n)} \equiv q_{\alpha_n}$.
If one uses the same $\tau_n$ in the extrapolation step and in the calculation of $\widehat{\tau}_n'(\alpha_n)$, the composite extrapolating least asymmetrically weighted squares estimator is
\[
\widetilde{\xi}_{\widehat{\tau}_n'(\alpha_n)}^{\star} = \left( \frac{1-\widehat{\tau}_n'(\alpha_n)}{1-\tau_n} \right)^{-\widehat{\gamma}_n^{H}} \widetilde{\xi}_{\tau_n} = ((\widehat{\gamma}_n^{H})^{-1}-1)^{\widehat{\gamma}_n^{H}} \widetilde{\xi}_{\alpha_n}^{\star}.
\]
In other words, rewriting Equation~\eqref{eq:proportionality} as $q_{\alpha_n} \equiv \xi_{\tau'_n(\alpha_n)} \approx (\gamma^{-1}-1)^{\gamma} \xi_{\alpha_n}$, the composite extrapolating least asymmetrically weighted squares estimator can be constructed by inserting the estimator $\widehat{\gamma}_n^{H}$ and the extrapolating least asymmetrically weighted squares estimator at level $\alpha_n$ into the right-hand side of this approximation. This  
has not, to the best of our knowledge, previously been noted in the literature.  
When combining this with with our marginal expected shortfall estimators, this gives two estimators of $\mathrm{QMES}_{X,\alpha_n}$. The asymptotic properties of the estimators discussed so far are established next.
\begin{Theo}
\label{theo:asyext2stage}
Suppose the conditions of Theorem~\ref{theo:asyexttime} hold with $\alpha_n$ in place of $\tau'_n$. Then, if $\widehat{\gamma}_n = \widehat{\gamma}_n^H$ and $\overline{\xi}^{\star}$ is either $\widehat{\xi}^{\star}$ or $\widetilde{\xi}^{\star}$, we have 
\[
\frac{\sqrt{n(1-\tau_n)}}{\log[(1-\tau_n)/(1-\alpha_n)]} \left( \frac{\overline{\xi}_{\widehat{\tau}'_n(\alpha_n)}^{\star}}{q_{\alpha_n}} - 1 \right) \to \mathcal{N}\left( \frac{\lambda_1}{1-\rho}, \gamma^2 \left[ 1+2\displaystyle\sum_{t=1}^{\infty} R_t(1,1) \right] \right),
\]
in the sense of weak convergence as $n\to\infty$.
\end{Theo}
An analogous asymptotic normality result holds for the composite estimators $\widetilde{\mathrm{XMES}}_{X,\widehat{\tau}_n'(\alpha_n)}^{\star}$ and $\widehat{\mathrm{XMES}}_{X,\widehat{\tau}_n'(\alpha_n)}^{\star}$ of $\mathrm{QMES}_{X,\alpha_n}$. 
\begin{Theo}
\label{theo:asyext2stage_XMES}
Suppose the conditions of Corollary~\ref{coro:asyQMES} hold with $\alpha_n$ in place of $\tau'_n$. Then, if $\widehat{\gamma}_{X,n}=\widehat{\gamma}_{X,n}^H$ and $\overline{\mathrm{XMES}}^{\star} = \widetilde{\mathrm{XMES}}^{\star}$ (resp.~$\overline{\mathrm{XMES}}^{\star} = \widehat{\mathrm{XMES}}^{\star}$), then
\[
\frac{\sqrt{n(1-\tau_n)}}{\log[(1-\tau_n)/(1-\alpha_n)]} \left( \frac{\overline{\mathrm{XMES}}_{X,\widehat{\tau}'_n(\alpha_n)}^{\star}}{\mathrm{QMES}_{X,\alpha_n}} -1 \right)\to \mathcal{N}\left( 0, \gamma_X^2 \left[ 1+2\displaystyle\sum_{t=1}^{\infty} R_{X,t}(1,1) \right] \right),
\]
in the sense of weak convergence as $n\to\infty$.
\end{Theo}

%
%
\section{Simulation experiments}
\label{sec:simulations_time_series} 
\subsection{Extreme expectile estimation}
\label{sec:simulations_time_series_exp} 
Here we summarise a numerical study of the finite-sample performance of the point and interval expectile estimators at extreme levels. More details and results on a similar study carried out for the estimators at the intermediate level may be found in Section B of the Supplementary Material. We consider simulations from
\begin{enumerate}[label=(\alph*)]
\item the AR($1$) model 
$
Y_{t+1} = 0.8 \, Y_{t}+\varepsilon_{t+1},
$
where the innovations $\varepsilon_t$ are independent and have a $t_3$ distribution;
\item the ARMA($1$,$1$) model
$
Y_{t+1} = 0.95 \, Y_{t}+\varepsilon_{t+1} + 0.9 \, \varepsilon_{t},
$
where the innovations $\varepsilon_t$ are independent and have a symmetric Pareto distribution with shape parameter $\zeta=3$; 
\item the ARCH($1$) model
$
Y_{t+1} = \sigma_{t+1} \varepsilon_{t+1},
$
where 
$
\sigma_{t+1}^2 = 0.4 + 0.6\,Y_{t}^2,
$
and the $\varepsilon_t$ are independent standard Gaussian innovations; and 
\item the GARCH($1$,$1$) model
$
Y_{t+1} = \sigma_{t+1} \varepsilon_{t+1},
$
where 
$
\sigma_{t+1}^2 = 0.1 + 0.4\,Y_{t}^2 + 0.4\,\sigma_t^2,
$
and the $\varepsilon_t$ are independent standard Gaussian innovations. 
\end{enumerate}
The first two models have strong linear serial dependence, and the last two have quadratic serial dependence leading to heteroskedasticity. If standard linear time series have heavy-tailed innovations with balanced tails then the tail index of $(Y_t)$ is that of the innovations, i.e., $1/3$ in models (a) and (b). The marginal distribution in (c) and (d) is heavy-tailed, and the tail indices, calculated using Theorem~2.1 in~\cite{miksta2000}, equal $0.262$ and $0.239$ respectively.

We simulate $10^{4}$ samples of size $n=2500$ for each model and consider the extreme level $\tau_n'=0.9995 \approx 1-1/n$;  values of $\xi_{\tau'_n}$ are in the top part of Table B.5 in the Supplementary Material.  
We calculate the extrapolating least asymmetrically weighted squares and quantile-based estimators using the intermediate level $\tau_n=1-k/n$ for $k\in \{6,8,\ldots,700\}$ for each simulated dataset and obtain various 95\% confidence intervals based on~\eqref{eq:conf_int_weiss}.  The big- and small-block sequences are chosen as $r_n=\lfloor \log^2(n)\rfloor$ and $l_n=\lfloor C\log n\rfloor$, where $C$ is selected such that $l_n$ is greater than or equal to a lag after which the largest of the sample autocorrelations of the data from the time series $(X_t)$ and $(X_t^2)$ are smaller than $0.1$. Checking whether the confidence intervals contain the true expectile allows us to compute the empirical coverage probability. 

The top row of Figure~\ref{fig:noncovprob_expct_extreme_level_09995_short} shows that our  confidence intervals have much better coverages than those that assume the data to be independent, and the coverages for intervals based on least asymmetrically weighted squares outperform those for the quantile-based estimators over a wider range of the intermediate sequence $k$. 
Coverages for the heteroskedastic models are satisfactory for a shorter range of the intermediate sequence $k$ and the intervals for these models tend to be slightly permissive.
The bias-corrected versions behave well in heteroskedastic models and are slightly conservative for linear time series. Overall, according to a rule of thumb to select a $k$ that falls in $[50, 200]$, our proposed confidence intervals perform very well for such a value of $n$. Similar results at level $\tau'_n=0.9999>1-1/n$ are found in Figure~B.2 of the Supplementary Material.

\subsection{Extreme marginal expected shortfall estimation}
\label{sec:simulations_time_series_MES} 
In order to assess the finite-sample performance of the marginal expected shortfall estimators at extreme levels we simulate data from four models  in which the sequences of bivariate innovations are independent and identically distributed:
\begin{enumerate}[label=(\alph*)]\setcounter{enumi}{4}
\item $
X_{t+1} = 0.8 \, X_{t}+\varepsilon_{X,t+1}
$
and 
$
Y_{t+1} = 0.8 \, Y_{t}+\varepsilon_{Y,t+1}
$.
For any $t$, the innovation $\varepsilon_{X,t}$ is distributed as $Z \,\ind( Z>0) - (-Z)^{1/2}\,\ind(Z<0)$, 
 $Z$ and  $\varepsilon_{Y,t}$ have $t_3$ distributions, and the dependence of $(\varepsilon_{X,t},\varepsilon_{Y,t})$ is given by a Student-$t$ copula with correlation parameter $\rho=0.8$ and three degrees of freedom; 
\item $
X_{t+1} = 0.95 \, X_{t}+\varepsilon_{X,t+1} + 0.9 \, \varepsilon_{X,t},
$
and 
$
Y_{t+1} = 0.95 \, Y_{t}+\varepsilon_{Y,t+1} + 0.9 \, \varepsilon_{Y,t}
$.
For any $t$, the innovation $\varepsilon_{X,t}$ is distributed as
$
Z \,\ind( Z>0) - (-Z)^{1/2} \,\ind( Z<0),
$
where $Z$ and $\varepsilon_{Y,t}$ have symmetric Pareto distributions with shape parameter $\zeta=3$,  and $(\varepsilon_{X,t},\varepsilon_{Y,t})$ have dependence given by a Gumbel copula with parameter $\theta=2$; 
\item $
X_{t+1} = \sigma_{X,t+1} \varepsilon_{X,t+1}$ and
$
Y_{t+1} = \sigma_{Y,t+1} \varepsilon_{Y,t+1}$, 
with 
$
\sigma_{X,t+1}^2 = 0.4 + 0.6\,X_{t}^2$ 
and 
$
\sigma_{Y,t+1}^2 = 0.4 + 0.6\,Y_{t}^2
$.
Each $\varepsilon_{X,t}$ has density
$
h(z) = 0.5 \, \ind( -1<z\leq 0 ) + 0.5 \, e^{-z} \,\ind( z>0 )$, $\varepsilon_{Y,t}$ is standard Gaussian, and the dependence of $(\varepsilon_{X,t},\varepsilon_{Y,t})$ is as in (a); and 
\item $
X_{t+1} = \sigma_{X,t+1} \varepsilon_{X,t+1},
$
with
$
\sigma_{X,t+1}^2 = 0.1 + 0.4\,X_{t}^2 + 0.4\,\sigma_{X,t}^2,
$
and
$
Y_{t+1} = \sigma_{Y,t+1} \varepsilon_{Y,t+1},
$
with 
$
\sigma_{Y,t+1}^2 = 0.1 + 0.4\,Y_{t}^2 + 0.4\,\sigma_{Y,t}^2.
$
Each $\varepsilon_{X,t}$ has density
$
h(z) = 0.5 \, \ind( -1<z\leq 0 ) + 0.5 \, e^{-z} \,\ind( z>0 )
$, 
 $\varepsilon_{Y,t}$ is standard Gaussian, and the dependence of $(\varepsilon_{X,t},\varepsilon_{Y,t})$ is given by a Gumbel copula with parameter $\theta=5$.
\end{enumerate}
The $Y_t$ components of models (e), (f), (g) and (h) are distributed according to models (a), (b), (c) and (d) respectively, so the models considered here extend those of Section~\ref{sec:simulations_time_series_exp}.

We simulate $10^4$ samples of size $n=2500$ for each model, and consider estimating $\mathrm{QMES}(\alpha_n)$ at levels $\alpha_n$ such that $\tau'_n(\alpha_n)=0.9995$ and $0.9999$, using our composite estimators $\widetilde{\mathrm{XMES}}_{X,\widehat{\tau}'_n(\alpha_n)}^{\star}$ and $\widehat{\mathrm{XMES}}_{X,\widehat{\tau}'_n(\alpha_n)}^{\star}$; recall that $(1-\alpha_n)\approx (\gamma_Y^{-1}-1)\{1-\tau'_n(\alpha_n)\}$ and the construction of $\widehat{\tau}'_n(\alpha_n)$ in Section~\ref{sec:CI}. The true values of $\mathrm{QMES}(\alpha_n)\equiv \mathrm{XMES}\{\tau'_n(\alpha_n)\}$, found by simulation, are in the bottom part of Table B.5 in Supplementary Material. Estimation at this level $\alpha_n$ gives us an idea of the performance of our composite estimation method, in a problem comparable in difficulty to that of Section~\ref{sec:simulations_time_series_exp}, at least as far as the $Y$ component is concerned. The confidence intervals are constructed as in Section~\ref{sec:simulations_time_series_exp}, applied to the $X_t$ component. In the quantile-based composite estimator $\widehat{\mathrm{XMES}}_{X,\widehat{\tau}'_n(\alpha_n)}^{\star}$, the Hill estimators of $\gamma_X$ and $\gamma_Y$ use the same $k$ for simplicity.

\begin{landscape}
\begin{figure}
	\centering
	\includegraphics[width=0.24\textwidth, page=1]{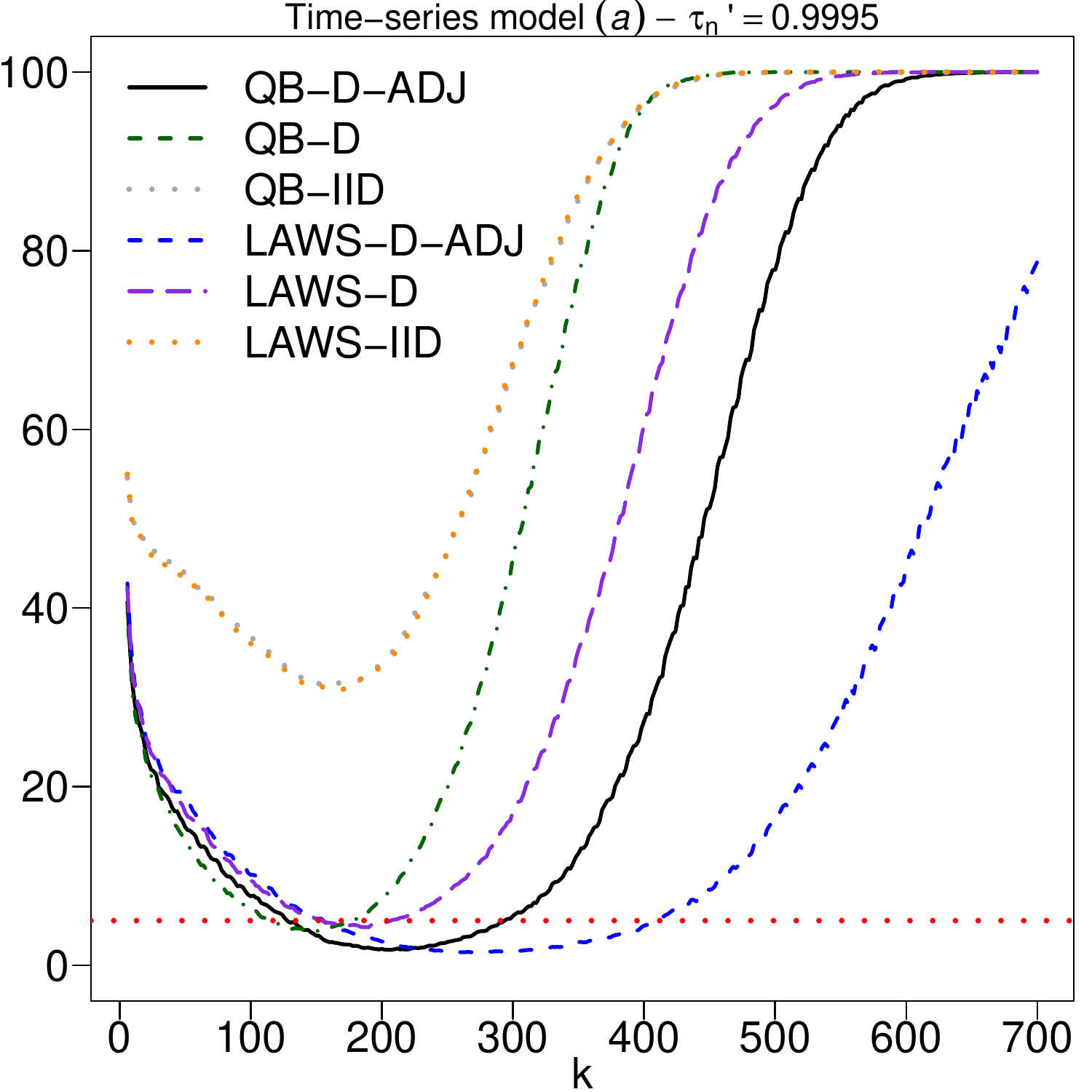}
	\includegraphics[width=0.24\textwidth, page=1]{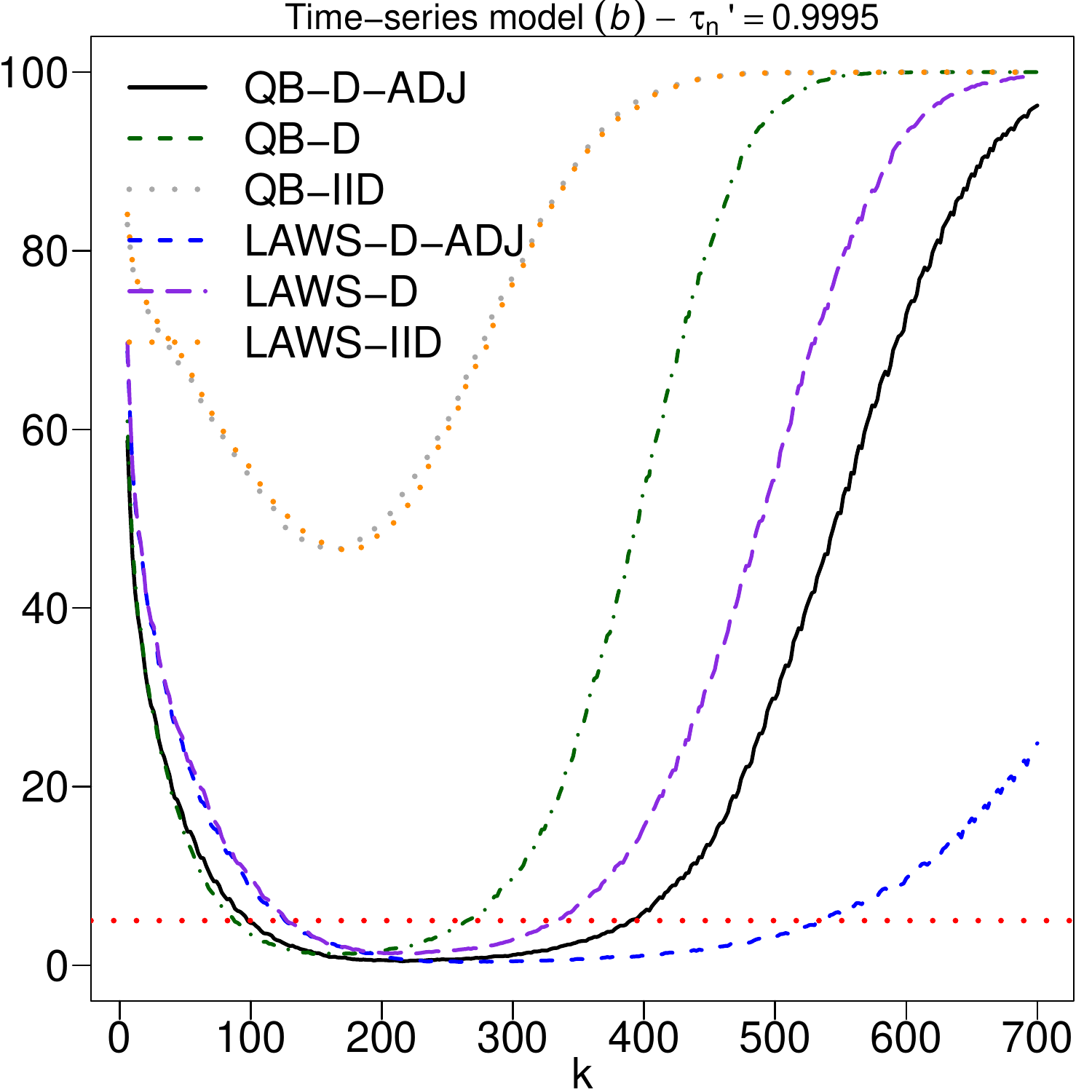}
	\includegraphics[width=0.24\textwidth, page=5]{PredExpctGarch.pdf}
	\includegraphics[width=0.24\textwidth, page=8]{PredExpctGarch.pdf}\\ 
	\includegraphics[width=0.24\textwidth, page=1]{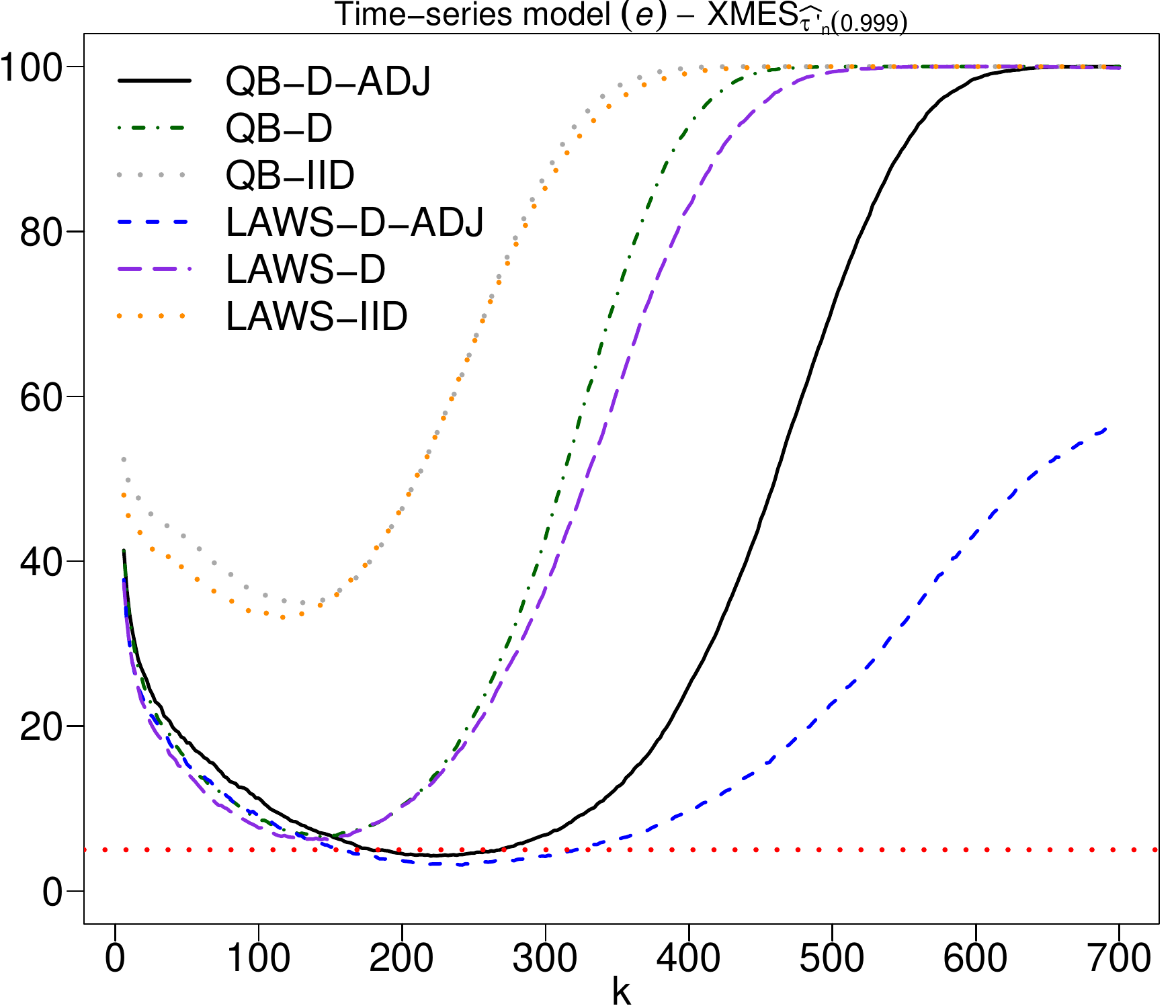}
	\includegraphics[width=0.24\textwidth, page=1]{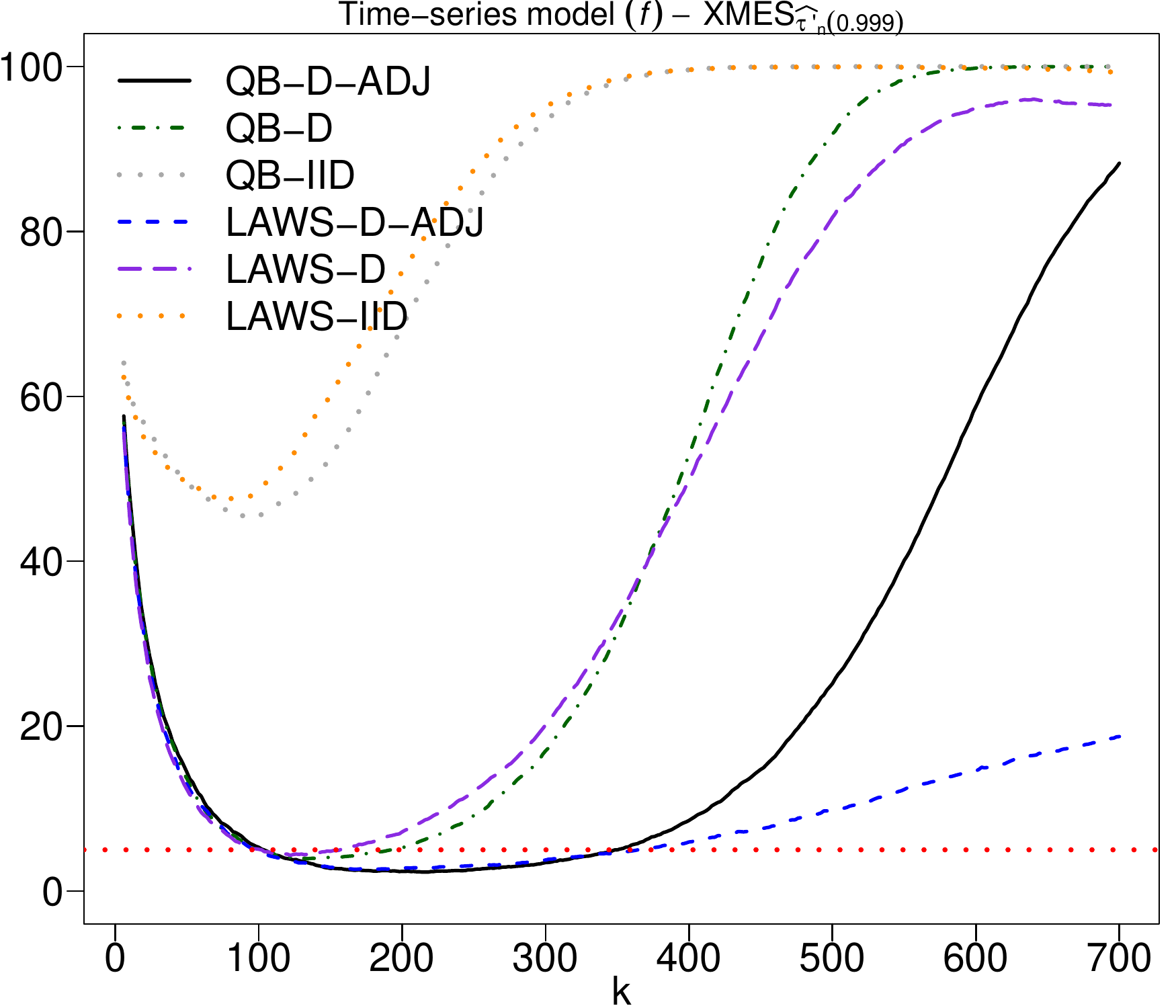}
	\includegraphics[width=0.24\textwidth, page=1]{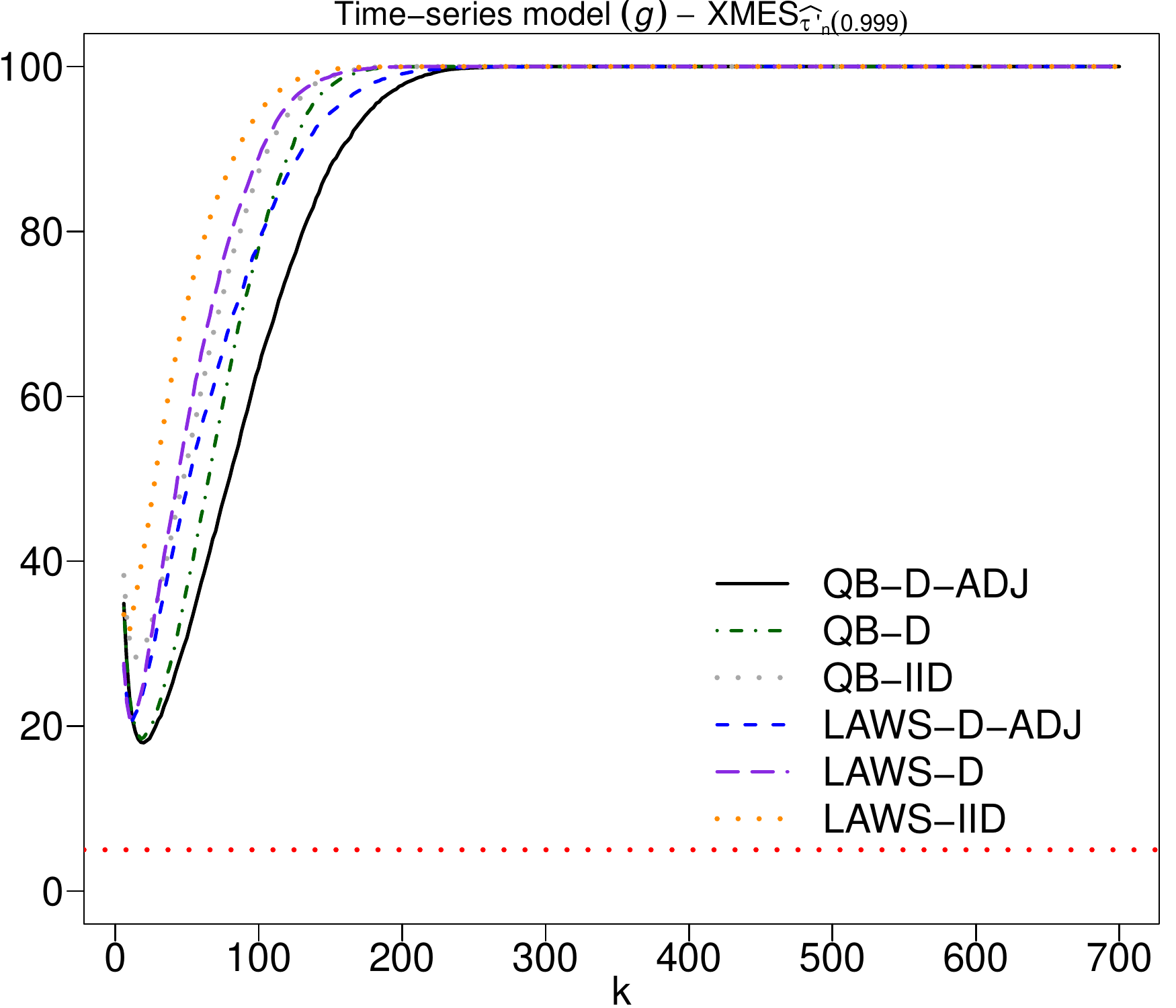}
	\includegraphics[width=0.24\textwidth, page=1]{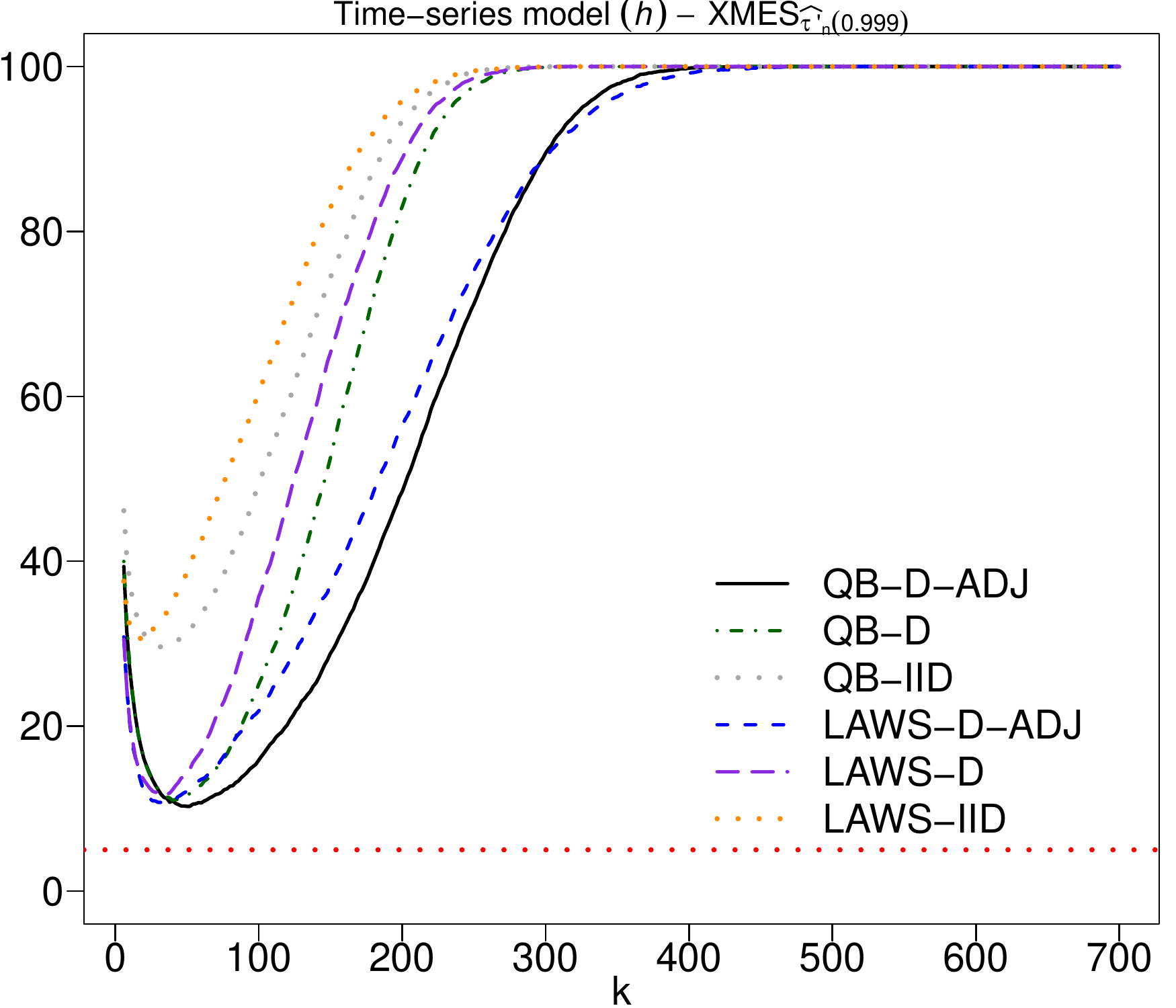}
\caption{Empirical error rate (\%) for nominal 95\% confidence intervals for the expectile $\xi_{\tau_n'}$ (top row, with $\tau_n'=0.9995$) and the marginal expected shortfall $\mathrm{QMES}(\alpha_n)$ (bottom row, with level $\alpha_n$ such that $\tau_n'(\alpha_n)=0.9995$). The true model is indicated above each panel.  The horizontal dotted red line represents the $5\%$ nominal error rate.  The quantile-based methods are QB-D-ADJ (using the extrapolating estimator), QB-D (setting $\widehat b=0$), QB-IID (ignoring the dependence), and likewise for the least asymmetric weighted square estimators.}
	\label{fig:noncovprob_expct_extreme_level_09995_short}
\end{figure}
\end{landscape}

The bottom row of Figure~\ref{fig:noncovprob_expct_extreme_level_09995_short} gives results at level $\tau'_n(\alpha_n)=0.9995$. The intervals for the linear time series have the same broad properties as in the upper panels.  
The coverages of our proposed confidence intervals are very good for linear time series, for a wide range of values of the intermediate sequence $k$ and for the bias-corrected and unadjusted versions.
The coverages with the heteroskedastic models (g) and (h) are less satisfactory, with our confidence intervals estimators providing reasonable results for a narrower range of values of $k$. Nevertheless, our intervals represent a significant improvement with respect to those relying on the theory for i.i.d.~observations. Also in this case, with the aforementioned rule of thumb (selecting $k\in [50, 200]$), our proposed confidence intervals provide reliable results overall.
Results at the level $\tau_n'(\alpha_n)=0.9999$ may be found in Figure B.3 of the Supplementary Material.

%
%
\section{Applications}
\label{sec:real_data_time_series} 
%
\subsection{Stock market index data}\label{market.sec}
\begin{landscape}
\begin{figure}
\centering
\includegraphics[width=0.24\textwidth, page=7]{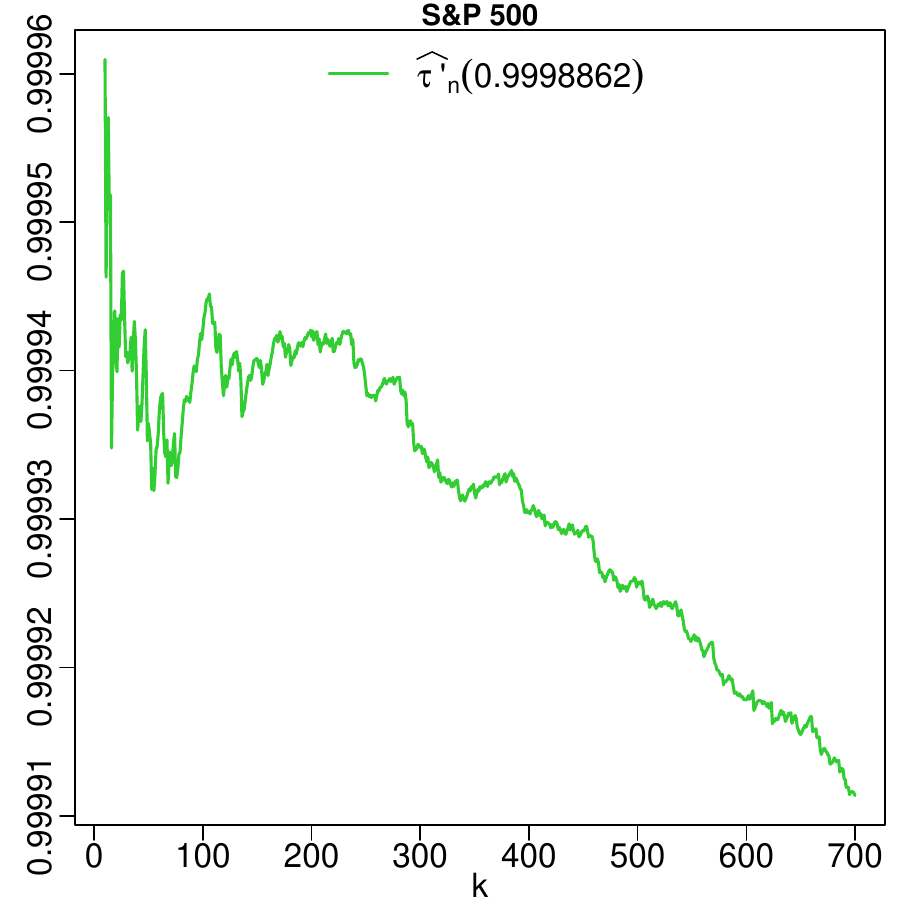}
\includegraphics[width=0.24\textwidth, page=4]{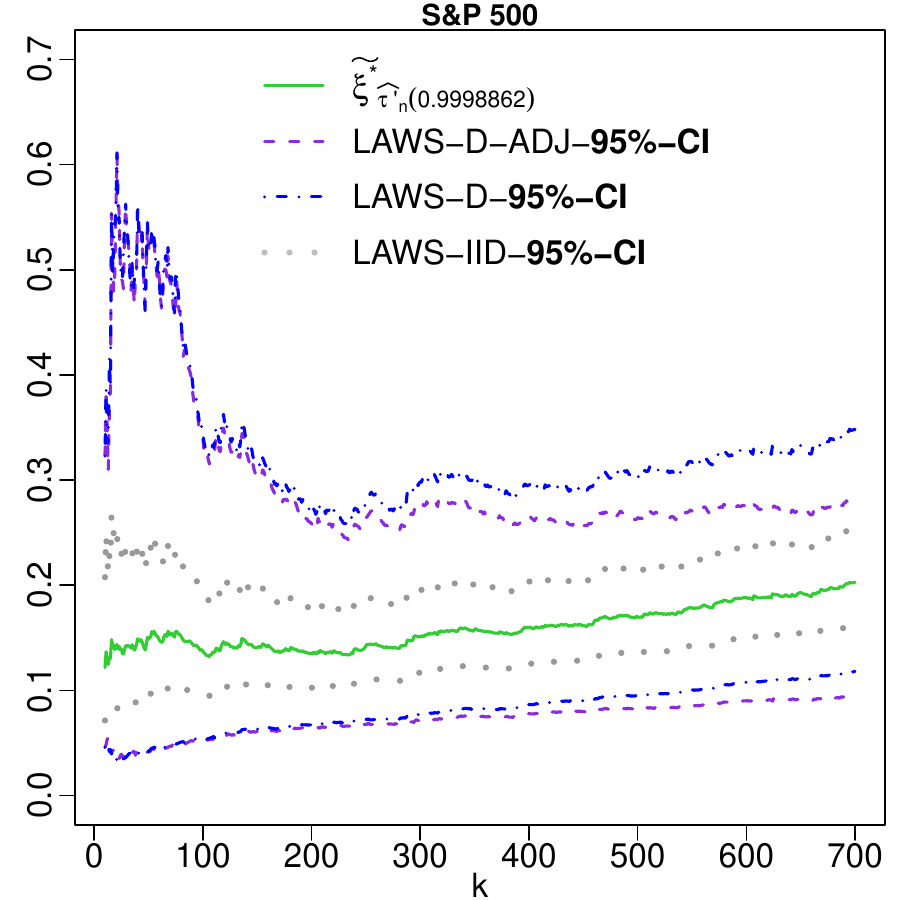}
\includegraphics[width=0.24\textwidth, page=5]{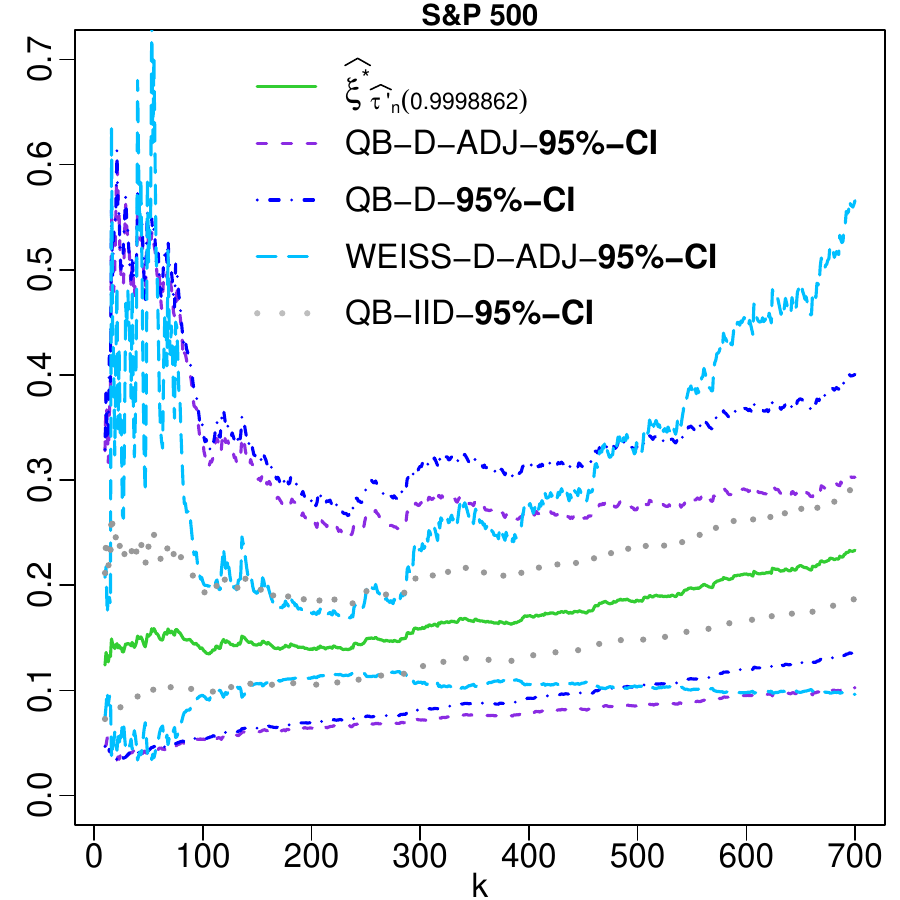}
\includegraphics[width=0.24\textwidth, page=11]{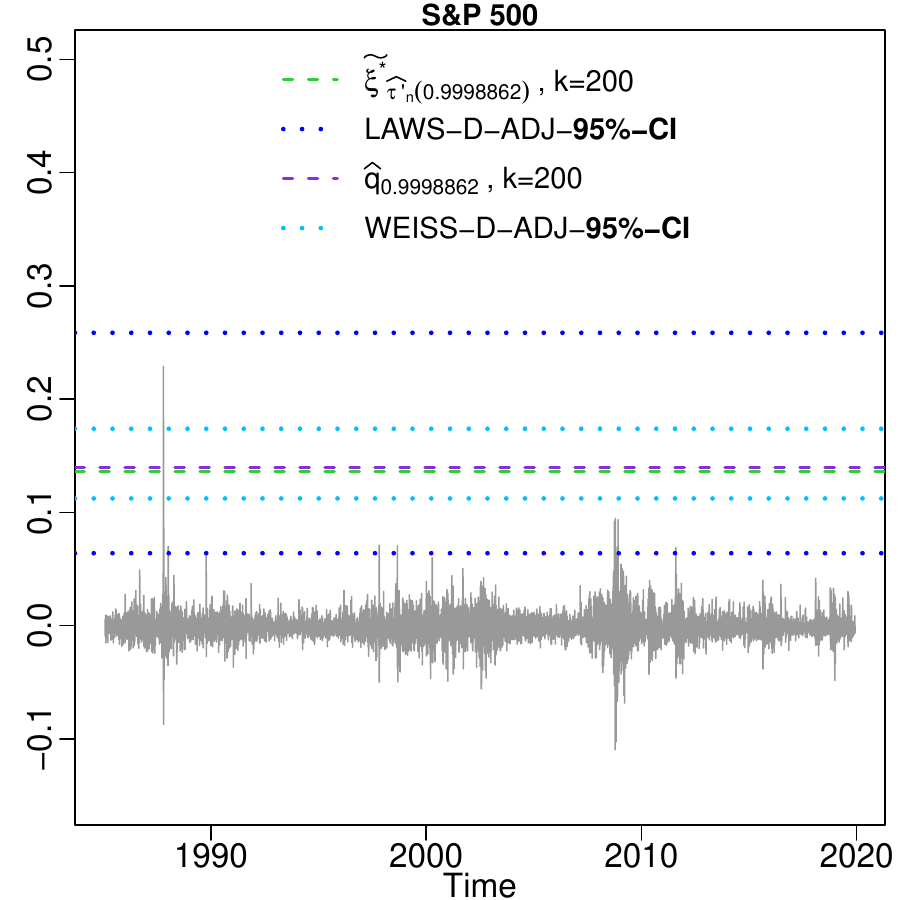} \\ 
\includegraphics[width=0.24\textwidth, page=7]{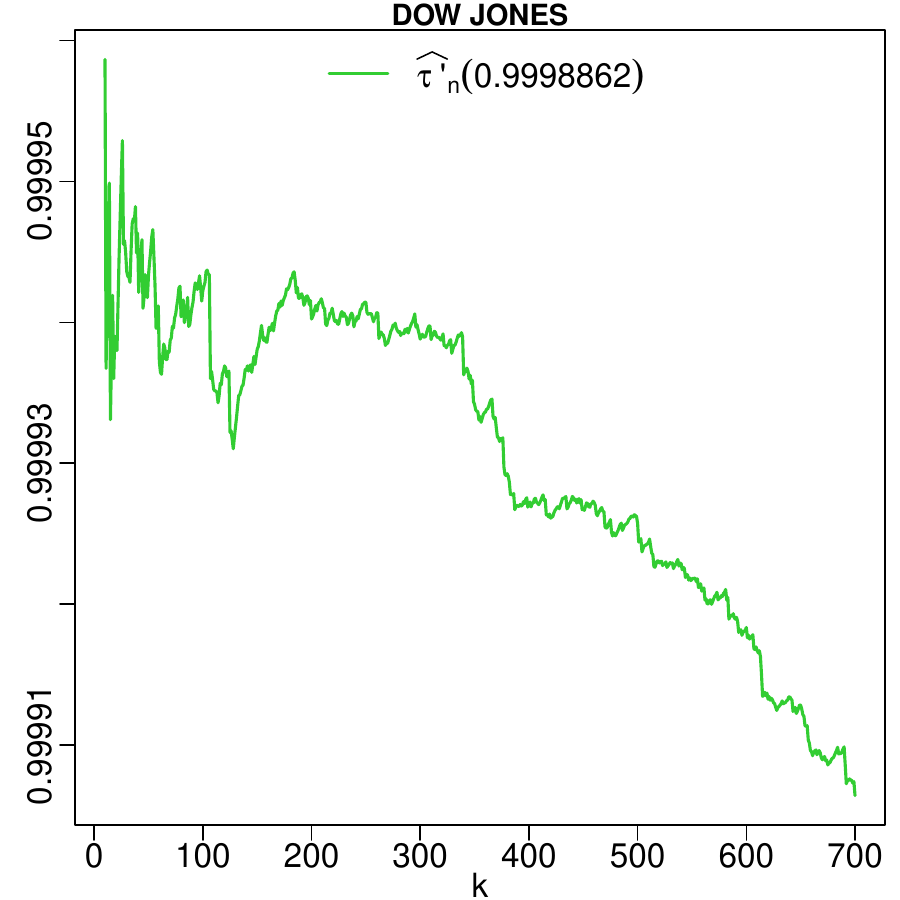}
\includegraphics[width=0.24\textwidth, page=4]{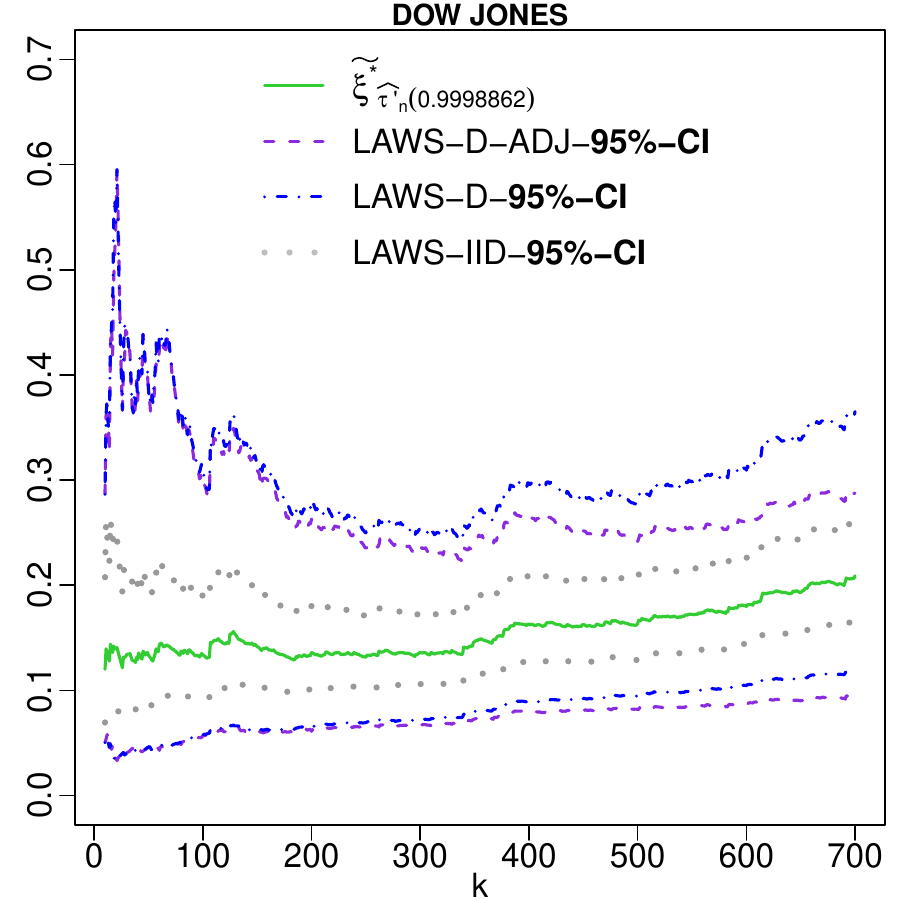}
\includegraphics[width=0.24\textwidth, page=5]{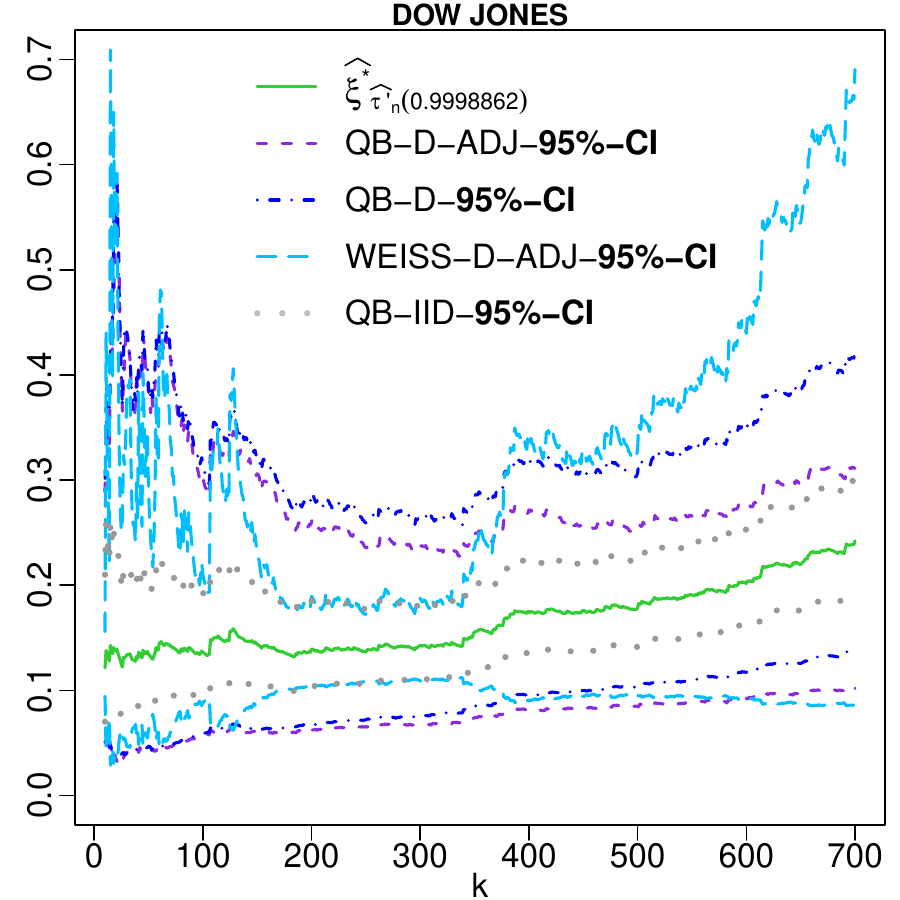}
\includegraphics[width=0.24\textwidth, page=11]{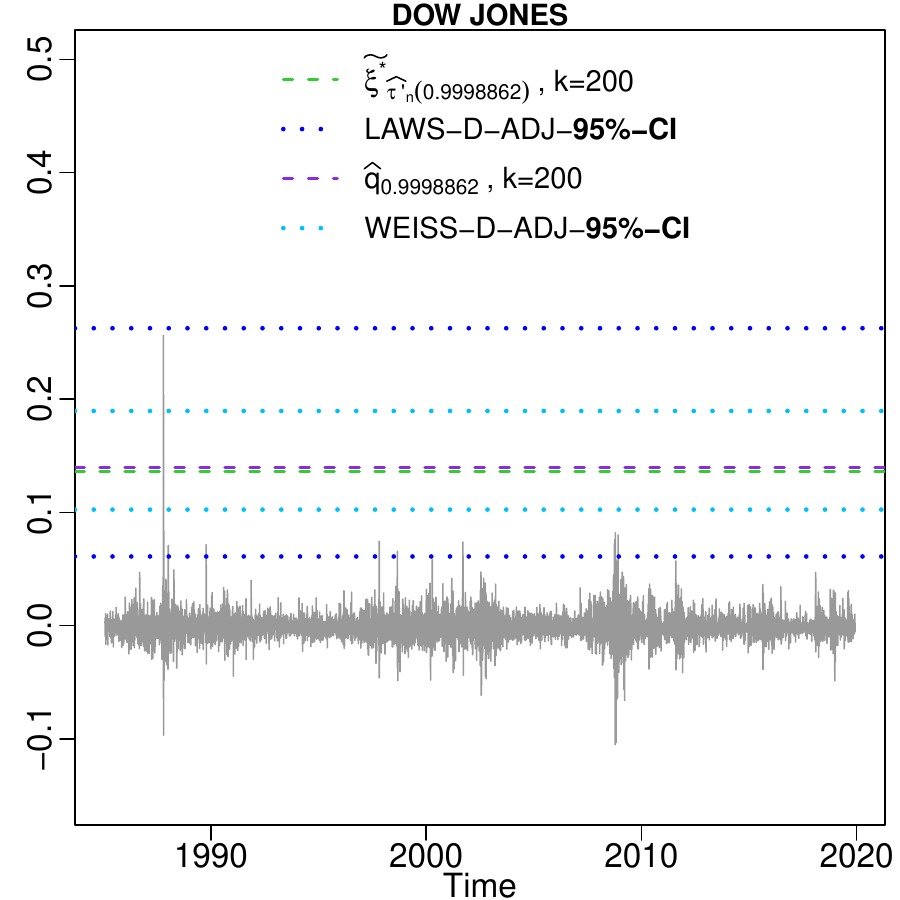}
	\caption{Financial data analysis. Analysis of S\&P 500 and Dow Jones industrial average data with $\alpha_n=0.9998862$. From left to right: estimate $\widehat{\tau}'_n(\alpha_n)$; composite extrapolated least asymmetrically weighted squares estimate at level $\widehat{\tau}'_n(\alpha_n)$, with $95\%$ LAWS-IID, LAWS-D and LAWS-D-ADJ confidence intervals; composite extrapolated quantile-based estimate at level $\alpha_n$, with $95\%$ QB-IID, QB-D, QB-D-ADJ and WEISS-D-ADJ confidence intervals; the data, with composite extrapolated least asymmetrically weighted squares estimate at level $\widehat{\tau}'_n(\alpha_n)$ and the Weissman quantile estimate $\widehat{q}_{\alpha_n}^{\star}$, both with $95\%$ confidence intervals.}
	\label{fig:nlogreturns_SPDJ}
\end{figure}
\end{landscape}

The rightmost panels of the top two rows in Figure~\ref{fig:nlogreturns_SPDJ} show $n=8{,}785$ daily negative log-returns of the S\&P 500 and Dow Jones Industrial Average  indices from 29~January 1985 to 12~December 2019. The data illustrate stylised facts such as the heteroskedasticity and fat tails of financial time series~\citep{embklumik1997}. For these two series, the Hill, maximum likelihood  and moment-based estimators of the tail index~\citep[e.g.,][Chapter 3]{haafer2006} are fairly stable for $k\in [100,300]$ and suggest that these series have heavy right tails with $\gamma\approx 0.35$ (see top panels of Figure B.4 in the Supplementary Material). Below we use the Hill estimator to construct our estimates and confidence intervals. 

The analysis of tail risk of loss returns is typically based on estimated Value-at-Risk at the $99.9\%$ level \citep[e.g.,][]{dre2003, haamerzho2016} or at some level $\alpha_n=1-p_n$, for $p_n\leq 1/n$. \cite{beldib2017} showed that such estimates yield capital requirements similar to expectile-based forecasts, if the level $\tau_n'$ of the expectile is chosen carefully, at a higher level than that of the Value-at-Risk. Here we fix $p_n=1/n$ and $\alpha_n=1-p_n=0.9998862$, and we estimate $\tau'_n(\alpha_n)$ by $\widehat{\tau}'_n(\alpha_n)$. Then we estimate the expectile at the extreme level $\widehat{\tau}'_n(\alpha_n)$ using the composite extrapolating least asymmetrically weighted squares estimator $\widetilde{\xi}^{\star}_{\tau'_n}$ and the corresponding quantile-based estimator $\widehat{\xi}^{\star}_{\tau'_n}$ for $\tau'_n=\widehat{\tau}'_n(\alpha_n)$. This produces estimators of $\xi_{\tau'_n(\alpha_n)}$, which is also the Value-at-Risk $q_{\alpha_n}=q_{1-1/n}$. The leftmost panels of the first and second row in Figure~\ref{fig:nlogreturns_SPDJ} plot $\widehat{\tau}'_n(\alpha_n)$ against $k$ for $k\leq 700$, where $\tau_n=1-k/n$ as before. These estimates fluctuate initially, then stabilise around a common value, and finally drift away due to the inclusion of data from the centre of the distribution; taking $k=200$ seems reasonable. We check this by calculating the composite extrapolating least asymmetrically weighted squares and composite extrapolating quantile-based estimators and the various confidence intervals of Section~\ref{sec:CI} at level $\tau'_n = \widehat{\tau}'_n(\alpha_n)$.  
The estimates and confidence intervals appear to be fairly stable provided $k$ is not too small, the choice $k=200$ again seems sensible. As expected, the confidence intervals that take the dependence into account are wider than those that do not. With $k=200$, we find $\widehat{\tau}'_n(\alpha_n)\approx 0.9999423$ for the S\&P 500 data, and $0.9999402$ for the Dow Jones data, both rather higher  than the original $\alpha_n=1-p_n=0.9998862$. 

Figure~\ref{fig:nlogreturns_SPDJ} also  compares our extrapolating least asymmetrically weighted squares and quantile-based estimators with the Weissman extreme quantile estimator at level $\alpha_n$, i.e., 
\begin{equation}
\label{WEISS.eqn}
\widehat{q}_{\alpha_n}^{\star} = \left( \frac{1-\alpha_n}{1-\tau_n} \right)^{-\widehat{\gamma}_n^H} \widehat{q}_{\tau_n} = \left( \frac{1-\alpha_n}{1-\tau_n} \right)^{-\widehat{\gamma}_n^H} X_{n-\lfloor n(1-\tau_n) \rfloor,n}. 
\end{equation}
Confidence intervals for the extreme quantile $q_{\alpha_n}$ can also be constructed using this estimator: here we use a method of~\cite{dre2003} for the estimation of the variance component $w(\gamma,\bfR)$ in Section~\ref{sec:CI}. Unlike our estimator, this method does not rely on a big-block/small-block argument, see Formula~(33) of~\cite{dre2003}. The estimates and 95\% confidence intervals are reported in Table~\ref{tab:final_estimates_SPDJ} and are shown in the rightmost panels of the top two rows in Figure~\ref{fig:nlogreturns_SPDJ}. The point estimates are reassuringly similar, but  the third and fourth panels suggest that the confidence intervals based on $\widehat{q}_{\alpha_n}^{\star} $ are generally much more volatile than the least asymmetrically weighted squares interval; moreover, in a neighbourhood of $k\approx 200$, they are very close to the intervals based on independent observations. For our selected $k=200$,  the fourth column of Figure~\ref{fig:nlogreturns_SPDJ} shows that  these intervals do not contain the largest sample value, despite estimating $q_{\alpha_n} = q_{1-1/n}$, whereas the least asymmetrically weighted squares intervals do contain the sample maximum.

\subsection{Financial returns of individual banks}

\begin{landscape}
\begin{figure}
\centering
\includegraphics[width=0.24\textwidth, page=7]{goldmansachs.pdf}
\includegraphics[width=0.24\textwidth, page=12]{goldmansachs.pdf}
\includegraphics[width=0.24\textwidth, page=11]{goldmansachs.pdf}
\includegraphics[width=0.24\textwidth]{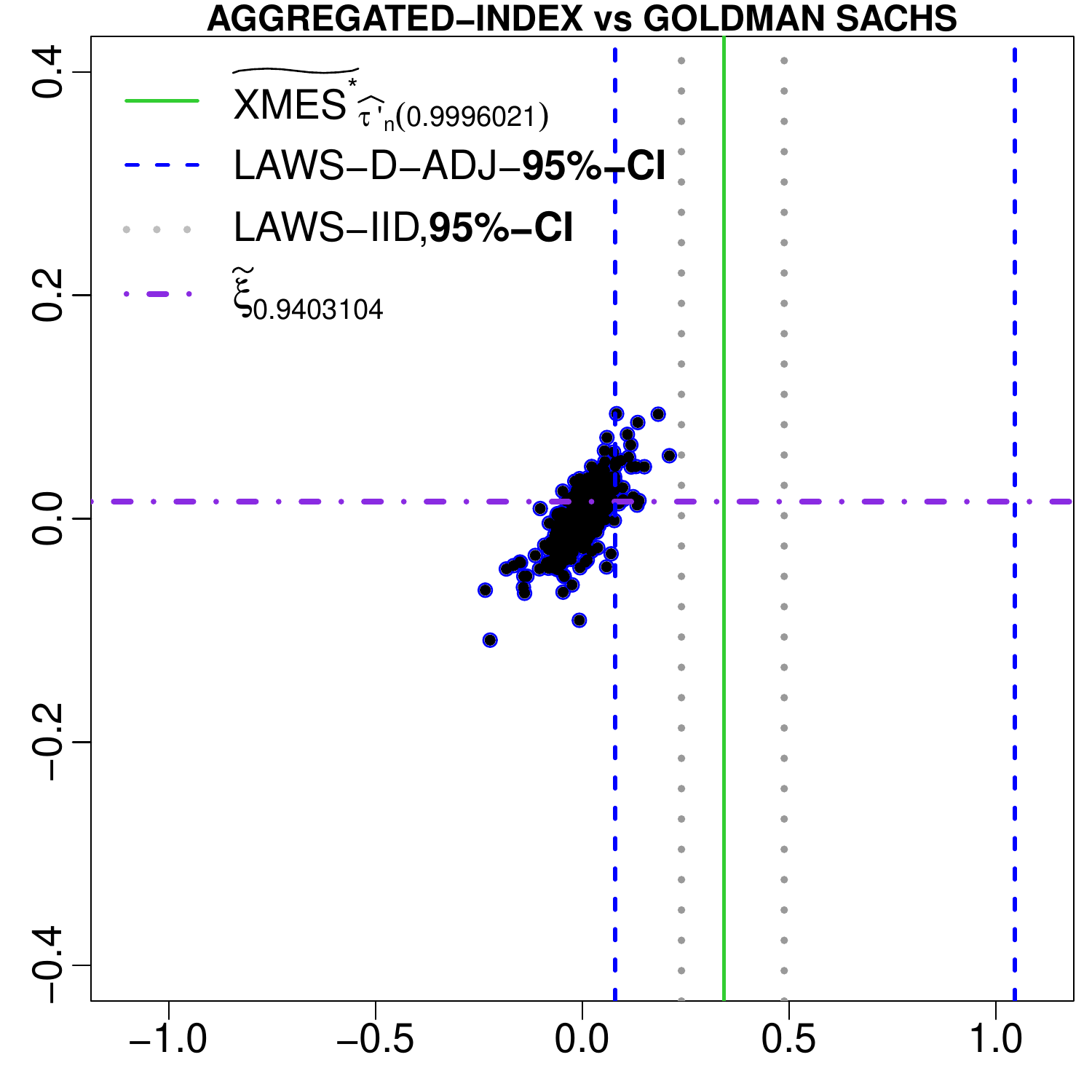} \\ 
\includegraphics[width=0.24\textwidth, page=7]{morganstanley.pdf}
\includegraphics[width=0.24\textwidth, page=12]{morganstanley.pdf}
\includegraphics[width=0.24\textwidth, page=11]{morganstanley.pdf}
\includegraphics[width=0.24\textwidth]{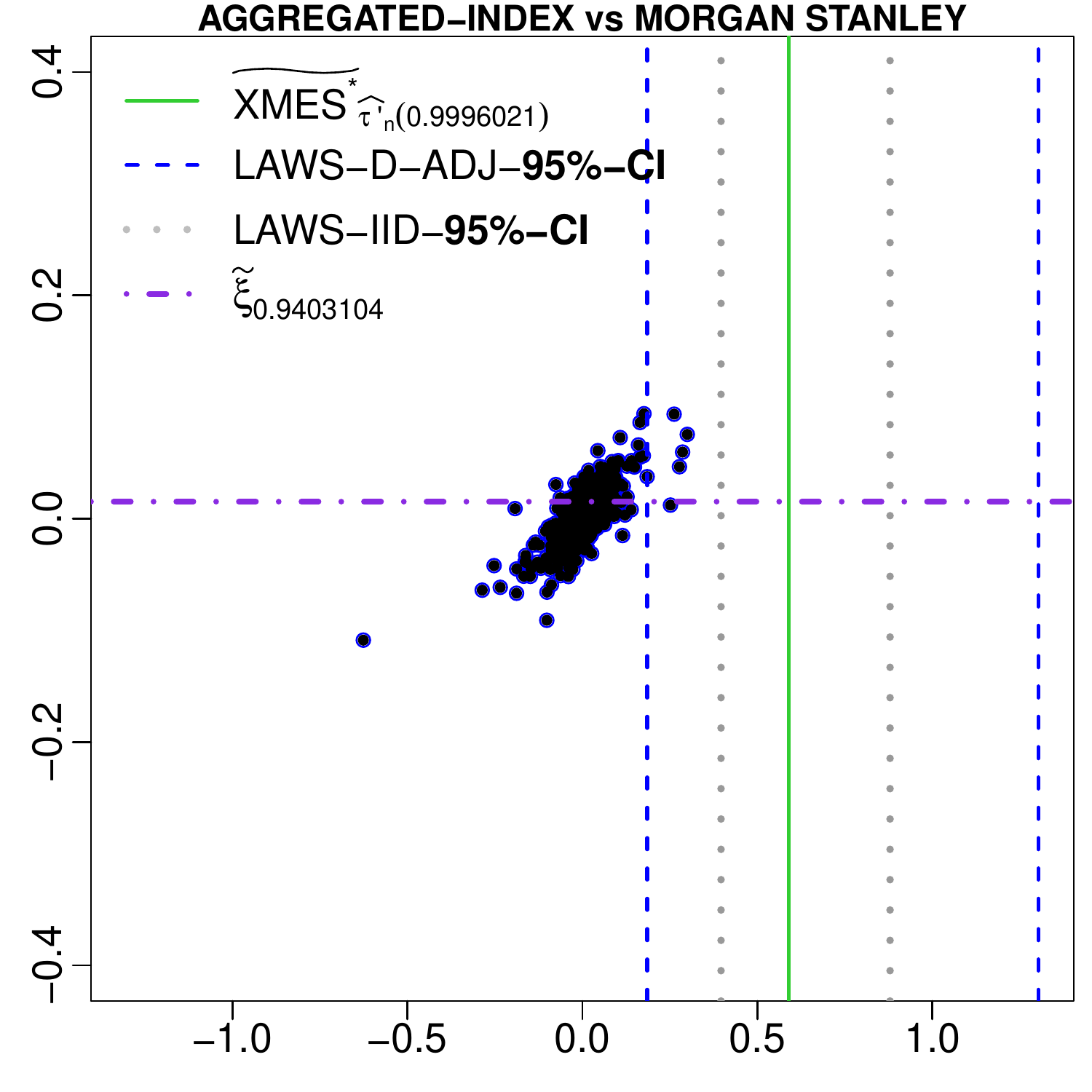}
	\caption{Financial data analysis. Analysis of Goldman Sachs and Morgan Stanley data with $\alpha_n=0.9996021$. From left to right: estimate $\widehat{\tau}'_n(\alpha_n)$; composite extrapolated least asymmetrically weighted squares estimate of\eqref{eq:truexmes} at level $\widehat{\tau}'_n(\alpha_n)$, with $95\%$ LAWS-IID, LAWS-D and LAWS-D-ADJ confidence intervals; least asymmetrically weighted squares estimate of~\eqref{eq:truexmes} at level $\alpha_n$, with $95\%$ QB-IID, QB-D, QB-D-ADJ and WEISS-D-ADJ confidence intervals; scatterplot of the data, with the composite extrapolated least asymmetrically weighted squares estimate of~\eqref{eq:truexmes} at level $\widehat{\tau}'_n(\alpha_n)$ with $95\%$ confidence intervals. The purple horizontal line is the intermediate level $\tau_n$.  The WEISS-D-ADJ confidence intervals are based on~\eqref{WEISS.eqn}; see the caption to Figure~\ref{fig:noncovprob_expct_extreme_level_09995_short} for the other  abbreviations.}
	\label{fig:nlogreturns_GSMS}
\end{figure}
\end{landscape}

We now analyse the financial returns of Goldman Sachs and Morgan Stanley in the context of systemic risk. We consider the daily negative log-returns $(X_t)$ on their equity prices from 3 July 2000 to 30 June 2010, alongside daily loss returns $(Y_t)$ of a value-weighted market index aggregating the New York Stock Exchange, the American Express Stock Exchange and the National Association of Securities Dealers Automated Quotation system for the same period. The corresponding tail index estimates again indicate heavy right tails (see bottom panels of Figure B.4 in the Supplementary Material, we have represented the same type of estimates for the $(Y_t)$ series in Figure B.5 of the Supplementary Material). The leftmost panels of the first and second row of Figure~\ref{fig:nlogreturns_SPDJ} display values of $\widehat{\tau}'_n(\alpha_n)$ against $k$ with again $\alpha_n=1-1/n =0.9996021$ and $\tau_n=1-k/n$.
%
%
%
%
%
\begin{table}[H]
\begin{center}
\begin{tabular}{lcc}
\toprule
Estimate & S\&P 500 & Dow Jones \\
\midrule
$\widehat{\gamma}_n^H$ & $0.336 \, [0.220,\,0.453]$ & $0.344 \, [0.222,\,0.467]$\\[5pt]
$\widetilde{\xi}^{\star}_{\widehat{\tau}_n'(\alpha_n)}$ & $0.136 \, [0.064,\,0.259]$ & $0.136 \, [0.061,\,0.263]$\\[5pt]
$\widehat{\xi}^{\star}_{\widehat{\tau}_n'(\alpha_n)}$ & $0.140 \, [0.064,\,0.258]$ & $0.139 \, [0.061,\,0.260]$\\[5pt]
$\widehat{q}_{\alpha_n}^{\star}$ & $0.140 \, [0.112,\,0.174]$ & $0.139 \, [0.103,\,0.190]$\\
\bottomrule
\end{tabular}
\caption{Estimates and 95\% confidence intervals for the daily log-returns for the S\&P 500 and Dow Jones  data with $k=200$ and $\alpha_n=1-1/n=0.9998862$.}
\label{tab:final_estimates_SPDJ}
\end{center}
\end{table}
\begin{table}[H]
\begin{center}
\begin{tabular}{lcc}
\toprule
Estimate & Goldman Sachs & Morgan Stanley \\
\midrule
$\widehat{\gamma}_{X,n}^H$ & $0.410 \, [0.172,\,0.648]$ & $0.459 \, [0.279	,\,0.639]$ \\[5pt]
$\widetilde{\text{XMES}}^{\star}_{X,\widehat{\tau}_n'(\alpha_n)}$ & $0.342 \, [0.079,\,1.046]$ & $0.590 \, [0.185,\,1.303]$ \\[5pt]
$\widehat{\text{XMES}}^{\star}_{X,\widehat{\tau}_n'(\alpha_n)}$ & $0.345 \, [0.080,\,1.055]$ & $0.603 \, [0.189,\,1.332]$\\[5pt]
$\widehat{\text{QMES}}^{\star}_{X,\alpha_n}$ & $0.345 \, [0.216,\,0.551]$ & $0.603 \, [0.294,\,1.236]$ \\
\bottomrule
\end{tabular}
\caption{Estimates and 95\% confidence intervals for the loss returns for Goldman Sachs and Morgan Stanley  data with $k=150$ and $\alpha_n=1-1/n=0.9996021$.}
\label{tab:final_estimates_GSMS}
\end{center}
\end{table}
These estimates slightly fluctuate initially, then stabilise around a common value, and finally drift away due to the inclusion of data from the centre of the distribution; taking $k=150$ seems reasonable.
With the same setup, the estimates obtained with the composite extrapolating least asymmetrically weighted squares and composite extrapolating quantile-based estimators of $\mathrm{QMES}_{X,\alpha_n}$, and the various confidence intervals at level $\tau'_n = \widehat{\tau}'_n(\alpha_n)$ are reported in first three panels of Figure \ref{fig:nlogreturns_GSMS}.
The estimates and confidence intervals are more stable for Goldman Sachs than for Morgan Stanley and for the least asymmetric squares method than for the quantile-based method.  
According to these results, taking $k=150$ seems reasonable, and we then find $\widehat{\tau}'_n(\alpha_n)\approx 0.9997239$ for the Goldman Sachs data and $0.9996626$ for the Morgan Stanley data, which we compare to the Weissman estimate $\widehat{\mathrm{QMES}}_{X,\alpha_n}^\star$. 
All estimates are reported in Table~\ref{tab:final_estimates_GSMS} and the least asymmetrically weighted squares estimates are shown in the rightmost panels of the bottom two rows of Figure~\ref{fig:nlogreturns_SPDJ}; they lead to conclusions similar to those in Section~\ref{market.sec}. For $k=150$ they give similar results on the Goldman Sachs data, and somewhat shorter confidence intervals on the Morgan Stanley data. Like on the stock market index data, the confidence intervals constructed with the method of~\cite{dre2003} are more volatile than ours.

In an application either to stock market index data or to an individual bank, it would be wise to inspect the analogue of Figure~\ref{fig:nlogreturns_SPDJ}, in order to identify stable regions for the estimates and confidence intervals.

\section*{Acknowledgements}
\noindent
G. Stupfler was previously based at the University of Nottingham, and the support of the Nottingham PEF Fund, of the French National Research Agency (grant ANR-19-CE40-0013) and of the AXA Research Fund is gratefully acknowledged. S.A. Padoan is supported by the Bocconi Institute for Data Science and Analytics. A. C. Davison is supported by the Swiss National Science Foundation.

\section*{Supplementary material}
\noindent
Supplementary material available at {\it Journal of Business \& Economic Statistics} online provides a detailed discussion of the technical conditions, further numerical details and results, and all proofs.

\bibliographystyle{agsm}

\bibliography{Bibliography}

\end{document}